\documentclass[aps,prl,twocolumn,groupedaddress,floatfix,superscriptaddress]{revtex4-1}

\usepackage{amsmath}
\usepackage{amssymb}
\usepackage{graphicx}
\usepackage{float}
\usepackage{bm}
\usepackage[english]{babel}
\usepackage{dsfont}
\usepackage{epstopdf}
\usepackage{textcmds}
\usepackage{soul}
\usepackage{color}
\usepackage{braket}
\usepackage[colorlinks=true,
            linkcolor=magenta,
            urlcolor=blue,
            citecolor=blue]{hyperref}

\usepackage[normalem]{ulem}

\begin{document}

\title{Strongly interacting photons in 2D waveguide QED}
\author{Matija Te\v{c}er}
\affiliation{Dipartimento di Fisica e Astronomia ``G. Galilei'', via Marzolo 8, I-35131 Padova, Italy}
\author{Marco Di Liberto}
\affiliation{Dipartimento di Fisica e Astronomia ``G. Galilei'', via Marzolo 8, I-35131 Padova, Italy}
\affiliation{Padua Quantum Technologies Research Center, Universit\'a degli Studi di Padova}
\affiliation{Istituto Nazionale di Fisica Nucleare (INFN), Sezione di Padova, I-35131 Padova, Italy}
\author{Pietro Silvi}
\affiliation{Dipartimento di Fisica e Astronomia ``G. Galilei'', via Marzolo 8, I-35131 Padova, Italy}
\affiliation{Padua Quantum Technologies Research Center, Universit\'a degli Studi di Padova}
\affiliation{Istituto Nazionale di Fisica Nucleare (INFN), Sezione di Padova, I-35131 Padova, Italy}
\author{Simone Montangero}
\affiliation{Dipartimento di Fisica e Astronomia ``G. Galilei'', via Marzolo 8, I-35131 Padova, Italy}
\affiliation{Padua Quantum Technologies Research Center, Universit\'a degli Studi di Padova}
\affiliation{Istituto Nazionale di Fisica Nucleare (INFN), Sezione di Padova, I-35131 Padova, Italy}
\author{Filippo Romanato}
\affiliation{Dipartimento di Fisica e Astronomia ``G. Galilei'', via Marzolo 8, I-35131 Padova, Italy}
\affiliation{Padua Quantum Technologies Research Center, Universit\'a degli Studi di Padova}
\affiliation{CNR-IOM Istituto Officina dei Materiali, Trieste, Italy}
\author{Giuseppe Calaj\'o}
\affiliation{Istituto Nazionale di Fisica Nucleare (INFN), Sezione di Padova, I-35131 Padova, Italy}

\date{\today}

\begin{abstract}

One dimensional confinement in waveguide Quantum Electrodynamics (QED) plays a crucial role to enhance light-matter interactions and to induce a strong quantum nonlinear optical response. In  two or higher dimensional settings, this response is reduced since photons can be emitted within a larger phase space,
opening the question whether strong photon-photon interaction can be still achieved. 
In this study, we positively answer this question for the case of a 2D square array of atoms coupled to the light confined into a two-dimensional waveguide. 
More specifically, we demonstrate the occurrence of long-lived two-photon repulsive and bound states with genuine 2D features.
Furthermore, we observe signatures of these effects also in free-space atomic arrays in the form of weakly-subradiant in-band scattering resonances.
Our findings provide a paradigmatic signature of the presence of strong photon-photon interactions in 2D waveguide QED.

\end{abstract}

\maketitle

The capability of engineering effective photon-photon interactions is a compelling requirement in many quantum technology applications for the generation and manipulation of complex states of light~\cite{chang2014quantum}. This task has been successfully achieved in light-matter interfaces, such as cavity QED~\cite{reiserer2015cavity}, Rydberg atomic ensembles \cite{firstenberg2016nonlinear,firstenberg2013attractive,liang2018observation} and waveguide QED~\cite{sheremet2023waveguide,roy2017colloquium}, where non-linear quantum optical effects are strongly enhanced. 
In particular, waveguide QED  has recently emerged as a versatile and experimentally implementable platform where the light confined in a 1D channel, either at optical~\cite{lodahl2015interfacing,hood2016atom,corzo2019waveguide,tiranov2023collective} or microwave frequencies \cite{astafiev2010resonance,brehm2021waveguide,mirhosseini2019cavity,kannan2023demand,shah2024stabilizing}, couples to one or multiple quantum emitters.
In this scenario, light confinement leads to strong photon correlations, resulting in anti-bunched~\cite{shen2007strongly,schrinski2022polariton}  or bunched output photons~\cite{shen2007strongly,shen2007stronglyL,zheng2011cavity,mahmoodian2018strongly,prasad2020correlating,mahmoodian2020dynamics,le2022dynamical}.
 Among the causes of the 
 bunching phenomena is the emergence of
multi-photon bound states~\cite{shen2007strongly,zheng2011cavity,mahmoodian2020dynamics,tomm2023photon} where two or more photonic excitations propagate jointly in the system with spatially correlated positions.
Remarkably, the lifetime and the dispersion properties of these states can be enhanced when ordered atomic arrays are coupled to the 
waveguide~\cite{zhang2020subradiant,poddubny2020quasiflat,bakkensen2021photonic,calajo2022emergence,PhysRevA.108.023707} or to waveguide networks~\cite{marques2021bound,schrinski2023photon}. In this setting, the realized bound state 
can be interpreted as a propagating excitation, which acts as a moving defect in the otherwise periodic 
medium,
dragging other photons with it.
The recent progress in building scalable microwave resonator arrays coupled to superconducting qubits~\cite{zhang2023superconducting,scigliuzzo2022controlling,gong2021quantum}, in interfacing 2D photonic crystals with cold~\cite{yu2019two} or artificial atoms~\cite{lodahl2015interfacing,hauff2022chiral,sipahigil2016integrated} and quantum emitters interacting with atomic matter waves~\cite{krinner2018spontaneous,kim2023super,PhysRevA.109.023306}
have created new experimentally viable opportunities for studying QED in 2D confined geometries.
This scenario is particularly fascinating  when combined with the possibility to induce photon-photon interactions 
and to build strongly correlated many-body quantum phases of light~\cite{RevModPhys.85.299,noh2016quantum}.

\begin{figure}[t!] 
\includegraphics[width=0.99\linewidth]{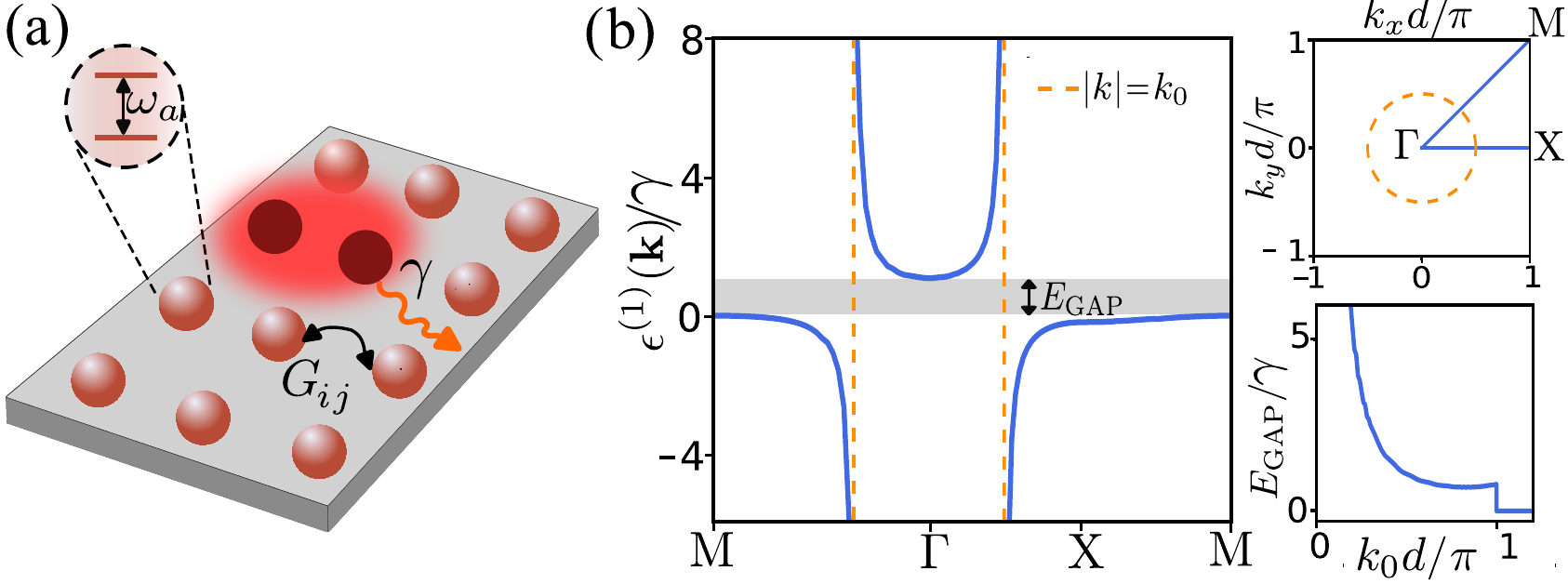}
    \caption{(a) A square array of two level atoms is coupled to the light confined into a 2D photonic waveguide. (b) Single photon dispersion relation for $k_0d=\frac{\pi}{2}$. Upper right panel: first Brillouin zone and $\Gamma,X,M$ symmetry points. Lower right panel: Bandgap energy as a function of the interatomic distance $k_0d$. The gap closes at $k_0d>\pi$. The dashed orange  lines indicate the dispersion's divergences  at $|\mathbf{k}|=k_0$. 
    }
    \label{fig:1}
\end{figure}

In this letter, we show that photonic states exhibiting both  spatial repulsion and  binding 
can occur by coupling a 2D atomic array to light confined into a 2D photonic structure (see Fig.~\ref{fig:1}). 
The emergence of such strong interactions in this 2D environment represents a nontrivial outcome. Indeed, unlike the 1D scenario, where photon-mediated interactions among emitters are infinite-range, in  2D, 
they decay as the square root of the emitters separation. 
We show that the interacting photonic states occurring in the system possess a distinct 2D structure, exhibiting binding and repulsion along different directions. Their existence 
is made possible 
 by an interference process originating from the 2D collective interaction of the emitters and does not rely on the engineering of the photonic bath~\cite{gonzalez2024light,RevModPhys.90.031002,PhysRevLett.124.213601,leonforte2024quantum}.
Furthermore, we  identify meta-stable in-band localized scattering resonances. Notably, these states also manifest in free-space atomic arrays, expanding the range of platforms where our findings can be  explored~\cite{rui2020subradiant,srakaew2023subwavelength,HUANG2023100470}.
Finally, we demonstrate how to excite the photon bound states via a dynamical relaxation process.
Our results provide a realistic pathway for realizing strongly-correlated states of light in 2D confined geometries in the sub-wavelength regime.

\textit{Model.} We consider a two-dimensional square array of $N$ two-level atoms  with  ground and excited states $|g\rangle$, $|e\rangle$ and lattice constant $d$ 
perfectly coupled~\cite{SuppMat} to the light confined into a 2D waveguide, as shown in Fig.~\ref{fig:1}. 
To simplify the description of this complex light-matter interacting system we integrate out the photonic degrees of freedom employing a Born-Markov approximation. 
This procedure is valid as long as the atom-photon dynamics, which is set by the atomic decay rate $\gamma$, occurs on a time-scale slower than a photon freely propagating through the whole array.
We obtain an effective spin model for the atoms described by the non-hermitian Hamiltonian ($\hbar = 1 
$)~\cite{Carmichael1999,Breuer2007,asenjo2017exponential,sheremet2023waveguide}
\begin{equation}\label{eq:Heff}
    \hat H_{\rm eff}=\sum_{ij}G\left(k_{0}|\mathbf x_i - \mathbf x_j |\right)\hat\sigma_{+}^{i}\hat\sigma_{-}^{j} ,\end{equation}
where 
$k_{0}=2\pi/\lambda$ is the photon wavevector whose corresponding frequency is resonant with the atomic transition frequency $\omega_a$ and $\hat\sigma_{+}^i=|e\rangle \langle g|$ and $ \hat\sigma_{-}^i=(\hat\sigma_{+}^i)^\dagger$
are the pseudospin operators for the $i$-th atom located at the position $\mathbf{x}_i$.  
The  dissipative and coherent long-range photon-mediated interactions among the atoms are encoded  in the function $G\left(k_{0}|\mathbf x_i - \mathbf x_j |\right)$, related to 2D electromagnetic Green's function  of the nanophotonic structure~\cite{asenjo2017exponential}. 
For confined photons with a quadratic isotropic dispersion relation, 
it reads \mbox{$G(z)=(\gamma/2)(\mathcal{Y}_0(z)-i\mathcal{J}_0(z))$}~\cite{SuppMat,gonzalez2015subwavelength,gonzalez2017markovian,galve2018coherent,gonzalez2018anisotropic}.  Here $\mathcal{J}_0(z)$ and $\mathcal{Y}_0(z)$ are respectively the zeroth order Bessel functions of the first and second kind, which decay as \mbox{$z^{-1/2}$.}
The Hamiltonian in Eq.~\eqref{eq:Heff} provides a complete description of the system within a fixed excitation sector in the absence of an external pumping field. 
For a single excitation, the Hamiltonian is diagonalized by Bloch waves, labelled by the  wavevector $\mathbf k$, having eigenvalues \mbox{$E^{(1)}(\mathbf k)=\epsilon^{(1)}(\mathbf k)-i\gamma^{(1)}(\mathbf k)/2$}, where $\epsilon^{(1)}(\mathbf k)$ and $\gamma^{(1)}(\mathbf k)$ provide the  single-excitation dispersion relation and collective decay rates, respectively  \cite{SuppMat}.
The dispersion relation (Fig.~\ref{fig:1}) exhibits two distinct polaritonic branches, along with a band gap near the atomic resonance frequency, where propagation of excitations is forbidden. 
Differently from  1D wQED~\cite{calajo2022emergence,sheremet2023waveguide}, the bandgap closes for inter-atomic distances above the sub-wavelength regime (see Fig.~\ref{fig:1}),  $k_0d>\pi$, namely for $d>\lambda/2$, due to the occurrence of higher diffraction orders. 
Note that, similar to  1D wQED ~\cite{albrecht2019subradiant,calajo2022emergence},  the divergences in the dispersion relation at the resonant wavevector, $|\mathbf{k}|=k_0$,  are associated with super-radiant modes~\cite{albrecht2019subradiant,zhang2019theory,kumlin2020nonexponential}. 

\begin{figure}[t!]
    \includegraphics[width=0.99\linewidth]{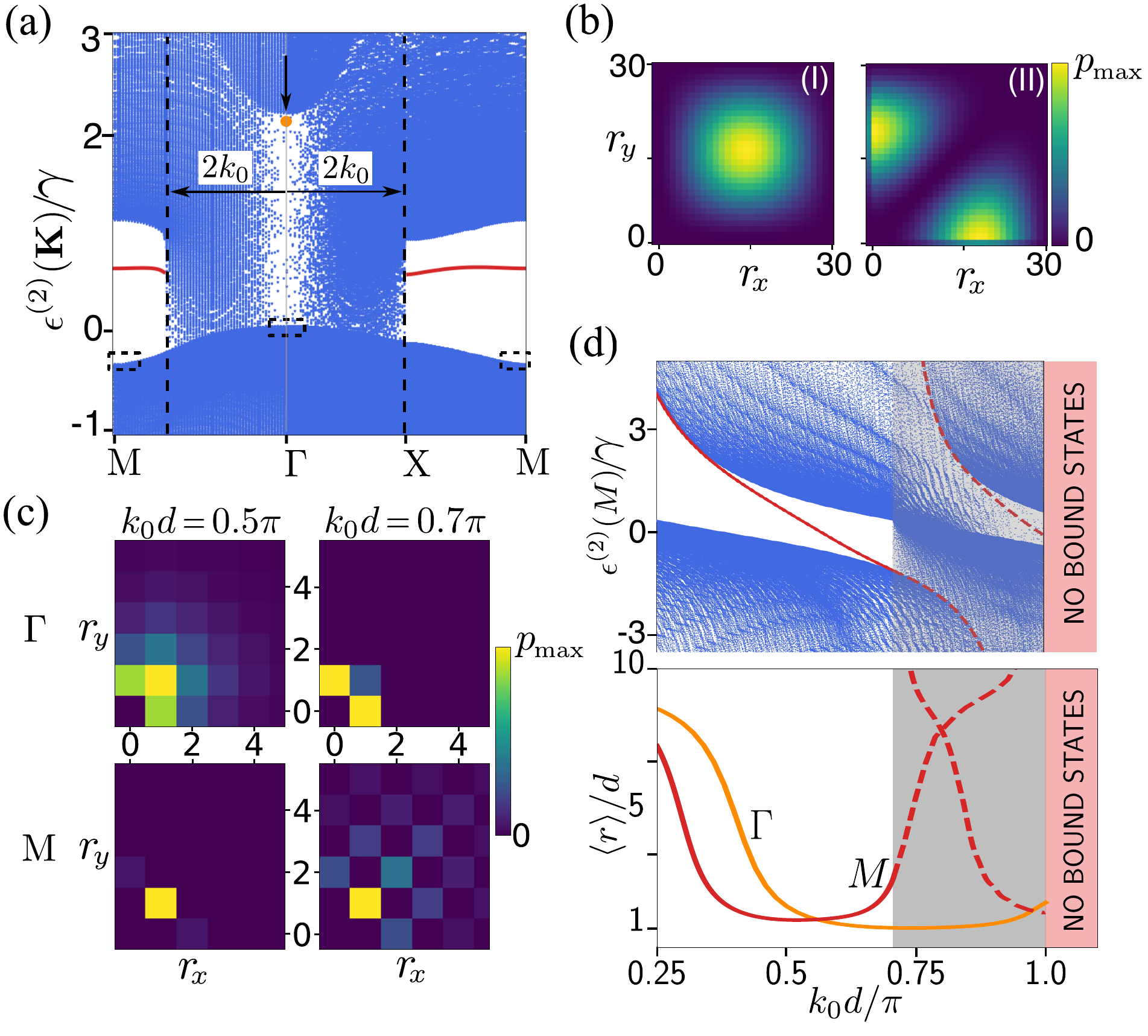}
    \caption{(a) Two-excitation spectrum as a function of the center of mass momentum for $k_0d=\frac{\pi}{2}$ . Blue dots represent unbound states  while the red lines and the orange dot represent the bound states. The dashed black boxes indicate the regions where  the repulsive states lie. (b) Relative coordinate population distribution, $p(r_x,r_y)$, of two kind of  repulsive states at $\Gamma$
    point for $k_0d=0.5$. Similar distributions up to a phase modulation are observed at the $M$ point.
    (c) Zoom of the bound states distribution at $\Gamma$ and M point for $k_0d=0.5\pi,0.7\pi$. (d)  Two-excitation spectrum at the $M$ point  (upper panel) and average relative distance between two excitations, $\langle r \rangle$, (lower panel) as a function of  $k_0d$.
     The  continuous lines indicate the BSs  while the dashed ones in the grey area indicate the scattering resonances branches. 
     }  \label{fig:2}
\end{figure}

\textit{Interacting photons.} To demonstrate that strong photon-photon interactions  occur in this 2D setting, we consider the two-body problem described by Eq.~\eqref{eq:Heff} restricted to the two excitations subspace. 
We focus on the thermodynamic limit of an infinite array, where there are no channels of dissipation and the spectra is real and determined by the coherent part of Hamiltonian~\eqref{eq:Heff}.
By re-parameterizing the arrangement of the atoms  in terms of center-of-mass ($\mathbf{R}$) and relative ($\mathbf{r}$) coordinates, we reduce the two-body problem to a single-particle one for the relative coordinate space and described by the basis set $|\mathbf K, \mathbf r\rangle$, where the parametric dependence on the center of mass momentum $\mathbf K$ is explicitly indicated~\cite{SuppMat}.
The corresponding Hamiltonian matrix elements read:

\begin{equation}\label{eq: HK}
(\hat{H}_\textrm{eff})^{\mathbf K}_{\mathbf r,\mathbf r'} 
    =\sum_{\epsilon=\pm}\cos (\mathbf{K}/2 \cdot \left(\mathbf{r}+\epsilon\mathbf{r}^{\prime}\right))\mathcal{R}\{G\left(k_0 \left|\mathbf{r}+\epsilon\mathbf{r}^{\prime}\right|\right)\}\,,
\end{equation}
where $\mathcal{R}$ denotes the real part. Numerical diagonalization returns 
the two-body spectrum, $\epsilon^{(2)}(\mathbf{K})$,  plotted in  Fig.~\ref{fig:2}(a). 
We observe a continuum of unbound two-particle states, whose energies correspond to the sum of two single-particle excitations, separated by a bandgap hosting a dispersive state. 
Most of the continuum states away from 
the high-symmetry points of the Brillouin zone (BZ)
are  scattering states made of a pair of particles with relatively high group velocity (see Fig.~\ref{fig:1}), thus experiencing a weak mutual interaction~\cite{sheremet2023waveguide}.
Instead, close to those high-symmetry points, the relatively flat single-particle dispersion can provide an enhancement of interactions.
It is convenient to map the matrix elements in Eq.~\eqref{eq: HK} onto a single-particle model for the relative coordinate $\mathbf r$ in the presence of an impurity potential, described by the Hamiltonian:
 \begin{equation}
    \label{eq: impurity H}
    \hat H^{(\rm imp)}_\mathbf{K}=\sum_{\mathbf r \mathbf r'}J^{\mathbf K}_{\mathbf r \mathbf r'}|\mathbf K, \mathbf r \rangle\langle \mathbf K, \mathbf r'|+U|\mathbf K, 0 \rangle\langle \mathbf K, 0|\,,
\end{equation}
where the first term describes free particle propagation, while the second term implements the short-range impurity potential with infinite strength, $U\rightarrow \infty$, located at $\mathbf{r}=0$. 
Here the hopping coefficient is \mbox{$J^{\mathbf K}_{\mathbf r \mathbf r'}=(1/N)\sum_{\mathbf{q}}\epsilon^{(2)}_{\text{scat}}(\mathbf{K},\mathbf q)e^{i(\mathbf r -\mathbf r')\mathbf{q}}$}, where $\mathbf{q}$ is the  relative momentum and the dispersion is given by the sum of individual photon energies, \mbox{$\epsilon^{(2)}_{\text{scat}}(\mathbf{K},\mathbf q)=\epsilon^{(1)}(\mathbf{q})+\epsilon^{(1)}(\mathbf{K}-\mathbf{q})$}. 
The effective model of Eq.~\eqref{eq: impurity H} is obtained by reformulating the Hamiltonian~\eqref{eq:Heff} in terms of hardcore bosons and then solving again the two-body problem in the relative coordinate frame~\cite{SuppMat}.
In this way, the  hardcore photon-photon interactions are mapped to the impurity term of Eq.~\eqref{eq: impurity H}, which reflects the saturation of an atom after the absorption of one excitation.

 
The first  consequence of strong photon interactions originating from the hard-wall condition induced by the impurity potential is the presence of two kinds of \emph{repulsive} scattering states indicated by the black dashed squares in Fig.~\ref{fig:2}(a).
Their population distribution in the relative coordinate 
shows a smooth repulsion of the two excitations either with respect to the $r_x=0$ and $r_y=0$ axis (type I) or with respect to the $r_x=r_y$ diagonal (type II), as shown in Fig.~\ref{fig:2}(b).
The 1D analogue of these states have been recently discussed both in free space~\cite{asenjo2017exponential} and wQED atomic arrays~\cite{zhang2019theory,albrecht2019subradiant,ostermann2019super,needham2019subradiance,zhong2020photon,kornovan2019extremely,Schrinski_polariton} and are commonly referred to as \qq{fermionic} states. 

Besides repulsive states we also identify the presence of  photon-photon bound states (BSs)~\cite{shen2007strongly,zheng2011cavity,mahmoodian2020dynamics,tomm2023photon,zhang2020subradiant,poddubny2020quasiflat,bakkensen2021photonic,calajo2022emergence}. Here the two excitations are spatially correlated and their occurrence can be understood, within the impurity  model~\eqref{eq: impurity H},  in terms of  a defect bound state localized  around $\mathbf{r}=\mathbf{0}$.
The BSs energy, $E_{\rm BS}(\mathbf{K})$, indicated by the red lines of Fig.~\ref{fig:2}(a), can be computed  by using  Green's function  methods~\cite{SuppMat,economou2006green} and it is given by the numerical solution of  $\mathcal{G}_0(0,E_{\rm BS}(\mathbf{K}))=1/U$ for $U\rightarrow\infty$, with 
 \begin{equation}
    \mathcal{G}_0(\mathbf r,E)=\frac{1}{(2\pi)^2}\int_{\rm BZ} d^2\bold{q}\frac{e^{i\bold{q}\cdot\bold{r}}}{E-\epsilon^{(2)}_{\text{scat}}(\mathbf{K},\mathbf q)}\,,
\end{equation}
being the free particle Green's function.
The BSs solutions need to respect the constraint $E_{\rm BS}(\mathbf{K})\neq \epsilon^{(2)}_{\text{scat}}(\mathbf{K},\mathbf q)$, meaning that their occurrence  is  conditioned to the existence of a band gap in the relative coordinate space. 
This requirement is not matched for $k_0d>\pi$ due to the closing of the single-excitation gap in 2D discussed before (see Fig.~\ref{fig:1}), which prevents the gap formation also for the two-particle spectrum. 
For $k_0d<\pi$, a gap in $\epsilon^{(2)}_{\text{scat}}(\mathbf{K})$ exists for values of the center of mass momentum in the range $|\mathbf{K}|>2k_0$ within the BZ. 
 This condition cannot be satisfied within the BZ  when $2k_0$ becomes equal to the $M$ point momentum, $|\mathbf{K}_M| = \sqrt 2 \pi/d$, and the gap  therefore closes~\cite{SuppMat}.
An additional gap occurs exactly at the $\Gamma$ point  due to a perfect vanishing of the density of states~\cite{bakkensen2021photonic,calajo2022emergence} and persists for all inter-atomic distances  $k_0d<\pi$.

\begin{figure}[t!]
    \includegraphics[width=0.99\linewidth]{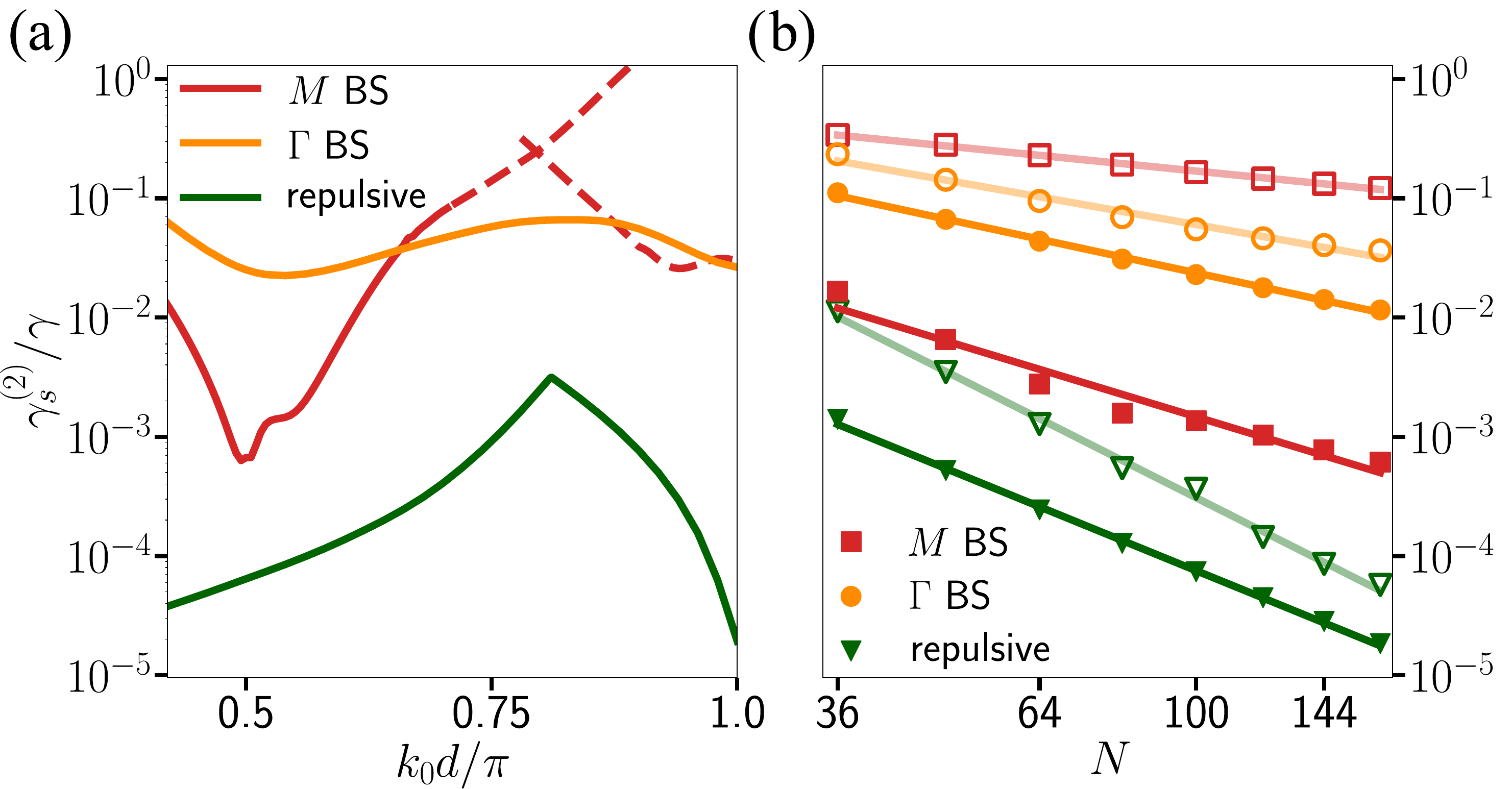}
    \caption{(a) Decay rates $\gamma^{(2)}_s$ of the interacting states as a function of  $k_0d$ for an array of $N=10\times10$ atoms. (b) Finite size scaling of the decay rates $\gamma_s$ with the system size $N$ in the 2D waveguide case (filled symbols) compared to the free space scenario (empty symbols). Here we fixed $k_0d=0.52\pi$ for the 2D waveguide  while $k_0d=1.09\pi$ and  $k_0d=0.73\pi$ for the $\Gamma$ and $M$ point in the free space case. In both plots, we considered as repulsive state the type II one. }  \label{fig:3}
\end{figure}
Once established the conditions of existence of the BSs,  we can use the impurity model to compute the bound states population distribution, $p(\mathbf{r})=|\mathcal{G}_0(\mathbf r,E_{\rm BS}(\mathbf{K}))|^2$, in the relative coordinate space. 
An example is shown in Fig.~\ref{fig:2}(c)  for the $\Gamma$ and $M$ points at two different inter-atomic distances. 
These plots clearly show a pronounced localization in the vicinity of $\mathbf{r}=0$, meaning that the two excitations are bound and spatially correlated. 
 We quantify this binding by the average distance between two excitations, $\langle r \rangle=\sum_{\mathbf{r}}p(\mathbf{r})|\mathbf{r}|$, which strongly depends on the inter-atomic distance, $k_0d$, as shown in Fig.~\ref{fig:2}(d).
 The plot shows how the bound states at $\Gamma$ and $M$ points are largely extended at short inter-atomic distances and progressively become more localized. 
 The $\Gamma$ point BS experiences minimal extension at the inter-atomic distance of $k_0d\sim 0.74\pi$, where a destructive interference process induced by the 
effective impurity is enhanced along the nearest neighbor direction~\cite{zhang2020subradiant}. When such an interference process is tailored along the next-nearest neighbor direction, the BS changes its relative coordinate distribution accordingly (see the first row of Fig.~\ref{fig:2}(c)). This occurs at $k_0d\approx 0.74/\sqrt{2}\pi \approx 0.52\pi$. The behavior of the $M$ point BS can be inferred from the center of mass  momentum induced cosine modulation in Eq.~\eqref{eq: HK}. Such modulation effectively creates a sub-lattice with  spacing $\sqrt{2}d$. The minima of the two curves in the lower panel of Fig.~\ref{fig:2}(d) are indeed displaced by such a factor, and the underlying modulation is visible in the bound state wavefunction shown in Fig.~\ref{fig:2}(c).
At $k_0d=\pi/\sqrt{2}$ the $M$ point gap closes and the BS ceases to exist. Nevertheless, we can still identify the presence of two localized  states in the band, which can be interpreted as scattering resonances,  states that contain both  unbound and  bound-state contributions~\cite{bakkensen2021photonic,PhysRevLett.126.083605,RevModPhys.66.539,SuppMat},
whose energies and localization are shown in Fig.~\ref{fig:2}(d). 
Interestingly, close to $k_0d=\pi$, we observe the formation of a region with low density of states (see Fig.~\ref{fig:2}(d)). This mechanism depends on the shape of the dispersion relation and  plays the role of an effective gap, for a finite size system, thus stabilizing them into quasi-bound states. 

\textit{Interacting states lifetime.} The  states described so far, being derived in the thermodynamic limit, have vanishing decay rates.
In  realistic settings, where the atomic lattice has a finite size, these states acquire a finite life-time due to leakage of the excitations at the array's edges. 
The decay rate of these states is provided by the imaginary part of the two-excitation sector eigenvalues, $\gamma^{(2)}_s$, obtained via exact diagonalization of Eq.~\eqref{eq:Heff} ~\cite{albrecht2019subradiant,yu2019two}, where $s$ labels the different states. 
For both repulsive and bound states, the decay rates are much smaller than the single atom emission rate, $\gamma$, indicating a sub-radiant behavior.
We find that, while  the $\Gamma$ point BSs have a weak dependence with the inter-atomic distance, the $M$ point BSs,  the associated scattering resonance  and the repulsive states strongly depend on that. For  both $\Gamma$ and $M$ point bound states the largest lifetime is  achieved around $k_0d\approx 0.52\pi$, resembling the localization plot shown in  Fig.~\ref{fig:2}(d)).
At this atomic distance, the decay rates of the $\Gamma$ and $M$ point BSs scale approximately as  $\sim N^{-1.5}$, as shown in Fig.~\ref{fig:3}.  The different  amplitude of the two BSs decay rates depends from a larger localization of the 
 $M$ photon bound state in the center of mass coordinate induced by a quasi-flat dispersion~\cite{SuppMat,poddubny2020quasiflat}.
The repulsive states exhibit an even stronger sub-radiant behaviour with a scaling of $\sim N^{-3}$,  which reflects the one of two independent single-excitation states ~\cite{SuppMat,asenjo2017exponential}.
These findings demonstrate that long-lived  photon interacting states  can occur even for finite size arrays. 

\textit{Free space array.}  
Some of the discussed interacting-states can be probed also in absence of a photonic
structure by considering  a 2D free space sub-wavelength atomic array.~\cite{bettles2016enhanced,shahmoon2017cooperative,manzoni2018optimization,PhysRevResearch.4.013110,PRXQuantum.2.040362,PhysRevA.104.033718}. 
By fixing  for simplicity the atomic polarisation to be orthogonal to the array's plane,  the projected dyadic Green's function ruling the atomic interactions in Eq.~\eqref{eq:Heff} reads~\cite{SuppMat,asenjo2017atom,asenjo2017exponential,lehmberg1970radiation,dung1998three}:
    $G(x)=(3\gamma_0)/(4(k_0x)^3)e^{ik_0x}\Big{(} 
     1-ik_0x-(k_0x)^2\Big{)}$,
where $\gamma_0=d_{eg}^2k_0^3/(3\pi\hbar\epsilon_0)$ is the free space spontaneous emission decay rate with $d_{eg}$ being the dipole moment strength.
In this scenario there is no bandgap in the single-excitation subspace and therefore neither in the two-excitations subspace. Thus, no BSs can form in this setting. 
However, 
it is still possible to identify  two-excitation scattering resonances exhibiting a localised wavefunction in their relative coordinate. 
For these states, we pinpoint regions characterized by a low density of states,  where they become metastable  within the sub-radiant region $k_0d<\sqrt{2}\pi$~\cite{SuppMat}.
This  is illustrated in Fig.~\ref{fig:3} by the scaling of sub-radiant decay rates at the $\Gamma$ and $M$ points, exhibiting dependencies of $N^{-1.2}$ and $N^{-0.7}$, respectively. It is noteworthy that these states can endure in the system for a timescale more than ten times greater than that associated with the decay of a single atom.
In addition to these localized states, we also identify the presence of repulsive states  presenting a similar decay rate scaling as the ones for a 2D waveguide. 
Note that, without the challenges related to the interfacing of the atoms to a photonic structure (see \cite{SuppMat}), these interacting excitations could potentially already be observed in current experimental implementations of subwavelength-ordered atomic arrays~\cite{rui2020subradiant,srakaew2023subwavelength,HUANG2023100470}.

\begin{figure}[t!]
    \includegraphics[width=0.99\linewidth]{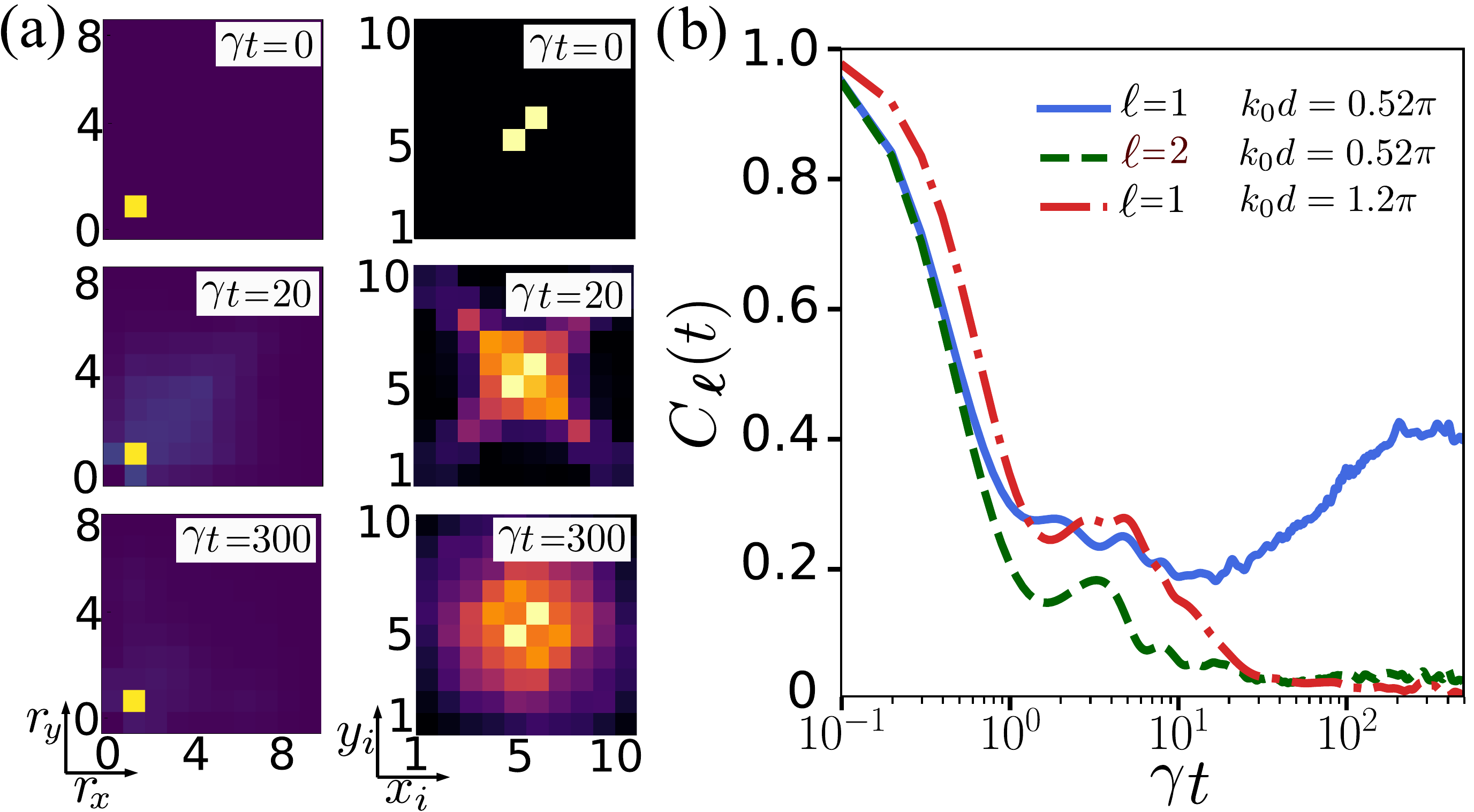}
    \caption{(a) Two-excitation probability distribution  in the  relative (left panel) and lab coordinate (right panel) space at times $\gamma t=0,20,300$ for $\ell=1$ and $k_0d=0.52\pi$. (b) Time-evolution of the two-particle correlator, $C_{\boldsymbol{\ell}}(t)$, for different inter-atomic distances $k_0d$ and different initial populations ($\ell=1,2$). 
   For all figures we fixed $N=10\times 10$.}  
    \label{fig:4}
\end{figure}

\textit{Dynamical excitation.} 
To observe the manifestation of BSs in the system dynamics, we consider 
a relaxation scenario where two atoms are initially  excited in a configuration that has a large overlap with the target state. Specifically, we consider the dynamical excitation of the $M$ point BS by exciting  two atoms in the bulk of the system separated by a distance $\mathbf{x}_i-\mathbf{x}_j=\boldsymbol{\ell}=(\ell,\ell)$, \mbox{$|\psi(t=0)\rangle=\hat\sigma_+^\mathbf{i}\hat\sigma_+^\mathbf{i+\boldsymbol{\ell}}|0\rangle$}. For an inter-atomic distance of $k_0d\approx 0.52\pi$ and an excitation separation of $\ell=1$, this configuration has indeed  a large initial overlap with the population distribution of the $M$ point BS~\cite{SuppMat}. 
In Fig.~\ref{fig:4} we plot the two excitation probability distribution in the relative and lab space coordinates at different times for this initial configuration.
Since the dominant contribution comes from the $M$ point BSs, the two excitations remain localized in  their relative position during the evolution, while  the BSs center of mass can diffuse across the lattice.
As a measurable observable to detect the dynamical excitation of the BS we define the two particle equal time correlator \mbox{$C_{\boldsymbol{\ell}}(t)=\sum_{\mathbf{i}}\langle \psi(t)| \hat{n}_\mathbf{i}\hat{n}_{\mathbf{i}+\boldsymbol{\ell}}|\psi(t)\rangle/|\langle \psi(t)| \psi(t)\rangle|^2$}, where the vector index $\mathbf{i}=(i_x,i_y)$ spans the whole array, $\hat{n}_\mathbf{i}=\hat\sigma_+^\mathbf{i}\hat\sigma_-^\mathbf{i}$ and the denominator normalizes this quantity with respect to the probability of having two excitations in the system.
This correlator indicates how likely is to find two excitations at a relative distance $\boldsymbol{\ell}=(\ell,\ell)$ in the entire array. Its time evolution is plotted in Fig.~\ref{fig:4} for different initial conditions. 
In particular, we consider the case $\ell=1$, the one having a large overlap with the BS, and $\ell=2$. 
In the first case, the two excitations stay bound together, as signaled by a persistent correlation at long times,
 whereas in the second case, the correlator quickly drops to zero. 
A large dropping is also observed,  for an inter-atomic distance of $k_0d\approx 1.2\pi$, where BSs do not exist.  Note that a dynamical procedure similar to the one presented for the BSs could be employed to excite scattering resonances either in a 2D waveguide or in free space.

\textit{Conclusions and Outlook.} 
We have discussed the emergence of two-photon strongly-correlated and long-lived states 
in a 2D array of atoms coupled to a two-dimensional waveguide. 
Using a spin model formulation, we have characterized these states discussing their dispersion, binding and decay properties and we have demonstrated their manifestation in the system dynamics.
The occurrence of these states crucially relies on interfacing an array of two-level atoms with the light confined into a two-dimensional photonic structure, but some of the discussed features could persist also in free space atomic arrays. 
This work opens interesting possibilities to study many-body physics with photons in this open and dissipative system.  In particular,
similarly as studied in 1D waveguide QED~\cite{mahmoodian2020dynamics,calajo2022emergence}, we could expect the existence of multi-photon bound states and a \qq{quantum to classical} transition toward the formation of 2D solitons~\cite{PhysRevLett.127.023603,PhysRevLett.127.023604}.
Targets of ongoing investigation include exploring  few-body interacting topological states~\cite{Ozawa2019, Salerno2018, Salerno2020, Clark2020, Leonard2023,PhysRevLett.126.103603,PRXQuantum.4.030306,PhysRevResearch.5.023031,PhysRevB.107.054301} in atomic arrays~\cite{perczel2017topological,perczel2017photonic,perczel2020topological} and emergent many-body phases in frustrated lattices \cite{bergholtz2013topological,Semeghini2021}. 
Another interesting perspective is to investigate the occurrence  of high dimensional bound states in 3D atomic arrays~\cite{brechtelsbauer2021quantum,andreoli2023maximum} where the atomic lattice completely fills the electromagnetic environment.

\emph{Acknowledgements.} This work was supported by the EU QuantERA2017 project QuantHEP, the EU QuantERA2021 project T-NiSQ, the  Quantum Technology Flagship project PASQuanS2, the NextGenerationEU project CN00000013, the Italian Research Center on HPC, Big Data and Quantum Computing, the  Quantum Computing and Simulation Center of Padova University, the INFN project QUANTUM, the Italian Ministry of University and Research via PRIN2022-PNRR project TANQU and the Rita Levi-Montalcini program and via the Departments of Excellence grant 2023-2027 “Quantum Frontiers” (FQ)
and by the project STRADA (Italian Presidency of the Council of Ministers).

\bibliography{refs}

\appendix
\clearpage

\onecolumngrid
\newpage

\pagebreak
\widetext

			\renewcommand{\theequation}{S\arabic{equation}}
			\renewcommand{\thefigure}{S\arabic{figure}}
			\renewcommand{\thepage}{S\arabic{page}}  
			\renewcommand{\thesection}{S\arabic{section}}
			\renewcommand{\thetable}{S\arabic{table}}
			\setcounter{equation}{0}
			\setcounter{figure}{0}
			\setcounter{page}{1}
			\setcounter{section}{0}

				\onecolumngrid 

\begin{center}
				Strongly interacting photons in 2D Waveguide QED \\ -- SUPPLEMENTARY INFORMATION --
\end{center}

\title{Interacting photons in 2D waveguide QED}
\author{M. Te\v{c}er}
\affiliation{Dipartimento di Fisica e Astronomia ``G. Galilei'', via Marzolo 8, I-35131 Padova, Italy}
\author{M. Di Liberto}
\affiliation{Dipartimento di Fisica e Astronomia ``G. Galilei'', via Marzolo 8, I-35131 Padova, Italy}
\affiliation{Padua Quantum Technologies Research Center, Universit\'a degli Studi di Padova}
\affiliation{Istituto Nazionale di Fisica Nucleare (INFN), Sezione di Padova, I-35131 Padova, Italy}
\author{P. Silvi}
\affiliation{Dipartimento di Fisica e Astronomia ``G. Galilei'', via Marzolo 8, I-35131 Padova, Italy}
\affiliation{Padua Quantum Technologies Research Center, Universit\'a degli Studi di Padova}
\affiliation{Istituto Nazionale di Fisica Nucleare (INFN), Sezione di Padova, I-35131 Padova, Italy}
\author{S. Montangero}
\affiliation{Dipartimento di Fisica e Astronomia ``G. Galilei'', via Marzolo 8, I-35131 Padova, Italy}
\affiliation{Padua Quantum Technologies Research Center, Universit\'a degli Studi di Padova}
\affiliation{Istituto Nazionale di Fisica Nucleare (INFN), Sezione di Padova, I-35131 Padova, Italy}
\author{F. Romanato}
\affiliation{Dipartimento di Fisica e Astronomia ``G. Galilei'', via Marzolo 8, I-35131 Padova, Italy}
\affiliation{Padua Quantum Technologies Research Center, Universit\'a degli Studi di Padova}
\affiliation{CNR-IOM Istituto Officina dei Materiali, Trieste, Italy}
\author{G. Calaj\'o}
\affiliation{Istituto Nazionale di Fisica Nucleare (INFN), Sezione di Padova, I-35131 Padova, Italy}

    

\maketitle

\section{2D Spin model derivation}\label{Sec.spinmodel}

Let us consider a 2D atomic ensemble interacting with the light confined into a 2D photonic structure, as illustrated in Figure 1 of the main text. To address the main physics originating from the 2D confinement we focus on the ideal waveguide QED scenario assuming a perfect coupling of the atoms to the 2D structure. With this assumption we  completely neglect spontaneous emission of the atoms into the free space. 
While this approximation may seem strong, it is frequently used in waveguide Quantum Electrodynamics (QED) literature~\cite{sheremet2023waveguide} and it accurately reflects the conditions encountered in scenarios involving superconducting qubits coupled to arrays of microwave resonators. In such cases, emission outside the resonators is typically two to three orders of magnitude smaller compared to emissions into other channels~\cite{scigliuzzo2022controlling}.
The Hamiltonian that describes this coupled atom-photonic structure system, under the rotating-wave and dipole approximations, is expressed as:
\begin{align}
    \label{eq: Full Hamiltionian}
    \begin{split}
        &\hat{H}=\hat{H}_0+\hat{H}_{\rm int}\,, \\
        &\hat{H}_0=\hbar \sum_{i}\omega_{a}\hat{\sigma}_{+}^i\hat{\sigma}_{-}^{i}+\hbar \int d^2{\mathbf{k}}\;\omega({\mathbf{k}})\hat{a}({\mathbf{k}})^{\dagger}\hat{a}({\mathbf{k}})\,, \\
 &\hat{H}_{\rm int}=\hbar\sum_{i}\int d^2\mathbf{k}\left(g_{\mathbf{k}}(\mathbf{r}_i)\hat{a}({\mathbf{k})}^{\dagger}\hat{\sigma}_{-}^{i}e^{i\mathbf{k}\cdot \mathbf{x}_i}+\text{h.c}. \right)\,,
    \end{split}
\end{align}
where  $\hat{a}(\mathbf{k})^{\dagger} (\hat{a}(\mathbf{k})$) are the bosonic creation (annihilation) operators of the photonic structure and 
\begin{equation}
    g_{\mathbf{k}}(\mathbf{r}_i)=\sqrt{\frac{\omega_{\mathbf{k}}} {2\hbar \epsilon_0 } }\mathbf{d}^{i}_{eg}\cdot \mathbf{\Phi}_{\mathbf{k}}(\mathbf{x}_i)\,,
\end{equation}
is the atom-photon coupling constant 
with $\epsilon_0$ being the vacuum permittivity, $\mathbf{d}^{i}_{eg}=\langle e|\hat{\mathbf{d}}^{i}|g\rangle$  the  dipole moment of the i-th atom and $\mathbf{\Phi}_{\mathbf{k}}(\mathbf{x}_i)$  the spatial mode eigenfunction of the field normalized according:
\begin{equation}
    \int d^2\mathbf{x}\mathbf{\Phi}_{\mathbf{k}}(\mathbf{x})\cdot\mathbf{\Phi}_{\mathbf{k'}}(\mathbf{x})=\delta_{\mathbf{k}\mathbf{k'}}.
\end{equation}
To simplify the description of this complex interacting system, we assume that the standard conditions for employing the Born-Markov approximation are satisfied. These conditions are guaranteed if the  density of states of the photonic bath is smooth and if  time retardation effects can be neglected. Consequently, we can eliminate the photonic degree of freedom, resulting in the following master equation for the atoms:~\cite{Carmichael1999,Breuer2007}:
\begin{equation}
    \label{eq: Master equation}
    \frac{d\hat{\rho}}{dt}=-\frac{i}{\hbar}\Big{[}\hat{H}_{\rm eff}\hat{\rho}-\hat{\rho} \hat{H}_{\rm eff}^{\dagger} \Big{]}+2\sum_{i,j} \Gamma_{ij} \hat{\sigma}_{-}^{i}\hat{\rho}\hat{\sigma}_{+}^j\,,
\end{equation}
where we have defined the effective non-hermitian Hamiltonian:
\begin{equation}
    \label{eq: Define general Heff}
    \hat{H}_{\rm eff}=\hbar\sum_{i,j}G_{ij}\hat{\sigma}_{+}^{i}\hat{\sigma}_{-}^j\,
\end{equation}
and the collective decay rate $\Gamma_{ij}=-\rm {Im}\{G_{ij}\}$.
Assuming a smooth position dependence of  the atomic coupling through the whole atomic ensemble $g_{\mathbf{k}}(\mathbf{r}_i)\simeq g_{\mathbf{k}}$ we finally obtain the following expression:
\begin{equation}
    \label{eq: interaction with integral}
    G_{ij}= \,\lim_{s \to 0^+}\int \frac{d^2\mathbf{k}}{2(2\pi)^2}\frac{|g_{\mathbf{k}}|^2}{is-(\omega_{a}-\omega_{\mathbf{k}})}e^{i\mathbf{k}\cdot \mathbf{x}_{ij}}, 
\end{equation}
where $\mathbf{x}_{ij}=\mathbf{x}_i-\mathbf{x}_j$\,. 
 Note that the master equation given in Eq.~\eqref{eq: Master equation}  for finite distances is generically multi-mode, being associated to multiple dissipation channels, and cannot be reduced to a single-mode description as in the Dicke model.

\begin{figure}
    \includegraphics[width=0.99\linewidth]{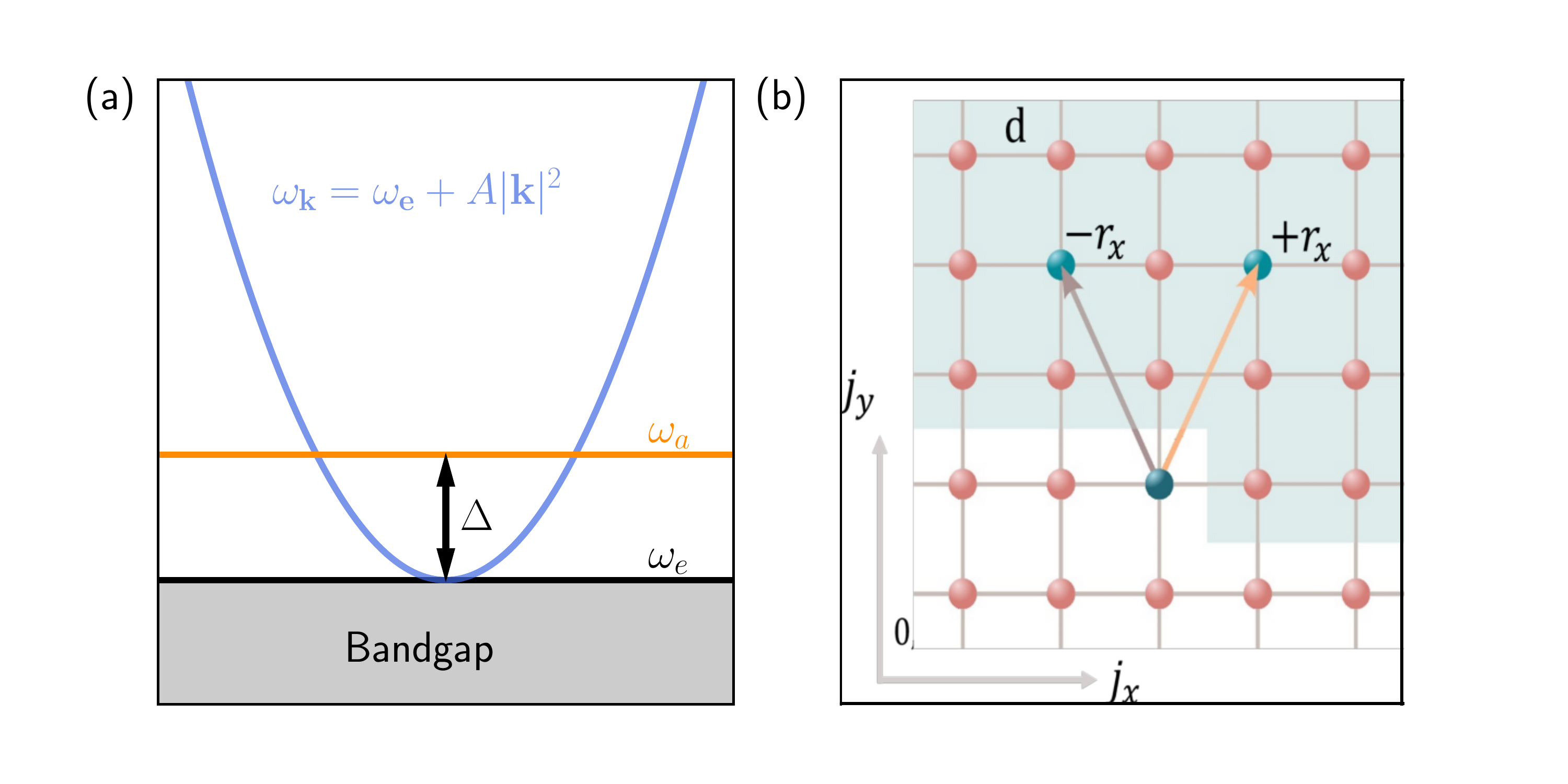}
    \caption{ (a)  The photonic structure's dispersion relation in the vicinity of a band edge is approximated to be quadratic and isotropic. The atomic transition frequency is located in the band ($\Delta=\omega_a-\omega_e>0$). (b) Schematic representation of the ansatz used to map the two-excitation sector into the relative coordinate frame. To avoid double counting, we consider the relative positions of the green highlighted atoms  only with respect to the atoms in the green shaded area.
    }  
    \label{fig:S1}
\end{figure}

\subsection{Isotropic and quadratic dispersion}
To simplify the discussion we  assume that the photonic dispersion is isotropic and quadratic. This choice is motivated by the common occurrence of such dispersion in 2D confined photonic structures like 2D photonic crystals~\cite{yu2019two} or above the band edge of a square photonic lattice~\cite{gonzalez2017markovian} (as illustrated in Fig.\ref{fig:S1}(a)).  It is worth noting that in the latter case the presence of saddle points in the middle of the band induces an anisotropic emission of the atoms. This behavior, being non-Markovian ~\cite{gonzalez2017markovian}, cannot be adequately captured by an effective spin Hamiltonian. Therefore, our model remains valid within this framework, as long as atomic frequencies are detuned from the center of the band.
Under these assumptions, the dispersion explicitly takes the form $\omega(\mathbf{k})=\omega_{e}+A|\mathbf{k}|^2$, where $\omega_e$ is the frequency at the band edge, and we consider $A>0$. If the atomic frequency $\omega_a$ lies inside the band, the atoms are coupled to propagating photons, and the collective interaction term is given by:
\begin{equation}
    G_{ij}= \lim_{s \to 0^+}\int_0^{k_c} \frac{k\,dk}{2(2\pi)^2}\frac{|g_{\mathbf{k}}|^2}{is-(\Delta -Ak^2)} \int_0^{2\pi}e^{i k |\mathbf{x}_{ij}| \cos \phi}d\phi=\lim_{s \to 0^+}\int_0^{k_c} \frac{k\,dk}{2(2\pi)}\frac{|g_k|^2}{is- (\Delta -Ak^2)}\mathcal{J}_0(k|\mathbf{x}_{ij}|)\,,
\end{equation}
where $\Delta=\omega_a-\omega_{e}>0$ is the atom-band edge detuning, $k_c$ sets an higher frequency cutoff and $\mathcal{J}_0(x)$ is the zeroth order Bessel function of the first kind.
Performing the change of variable $q=\sqrt{\frac{\Delta}{A}}k$ and assuming the coupling  to be constant across the integration domain, $g_{k}\approx g$, we finally obtain~\cite{gonzalez2015subwavelength}:
\begin{equation}
        G_{ij}\approx \lim_{s \to 0^+}\frac{ |g|^2}{A} \int_0^{\infty}\frac{q dq}{4\pi}\frac{\mathcal{J}_0\left(q|\mathbf{x}_{ij}|k_0\right)}{is-(1-q^2)}=
         \frac{\gamma}{2} \left(\mathcal{Y}_0\left(k_0|\mathbf{x}_{ij}|\right)-i\mathcal{J}_0\left(k_0|\mathbf{x}_{ij}|\right) \right)\, ,
\end{equation}
where $\mathcal{Y}_0(x)$ is the zeroth order Bessel function of the second kind, $\gamma=\frac{ |g|^2}{4A}$ is the single atom spontaneous emission decay rate, $k_0=\sqrt{\frac{\Delta}{A}}$ is the resonant wavevector  and we assumed $k_c\sqrt{\frac{A}{\Delta}}\gg 1$. 
In this form, $G_{ij}$ is explicitly composed of two contributions, accounting for coherent (first term) and dissipative (second term) collective interactions. Note that, similarly to free space arrays, the diagonal elements of the  coherent interaction diverge. Such divergence can be incorporated into a re-normalized resonance frequency and we practically set it to zero defining $G_{ii}=-i\gamma/2$.
The decay rate $\gamma$ establishes the scale of the system dynamics. Notably, the earlier assumption of neglecting time retardation effects can be formulated as $\gamma\ll c/L$, where $c=2Ak_0$ is the photon group velocity at the atomic frequency and $L$ is the size of the atomic ensemble.

\subsection{Single excitation dispersion}
In the thermodynamic limit, the eigenstates of the effective Hamiltonian~\eqref{eq: Define general Heff} with single excitations are spin waves characterized by well-defined quasimomentum $\mathbf{k}$. To diagonalize the Hamiltonian, a Bloch wave ansatz is employed, utilizing the Fourier transform of spin operators $\hat{\sigma}_{-}^{n}=\frac{1}{\sqrt{N}}\sum_{\mathbf{k}} e^{i\mathbf{k}\cdot \mathbf{x}_n}\hat{\sigma}_{-}^{\mathbf{k}}$ ($\hbar=1$):
\begin{equation}
    \label{eq: 1exc derivation 1}
\hat{H}_{\rm{eff}}=\sum_{n,m}G(k_0|\mathbf{x}_m-\mathbf{x}_n|)\hat{\sigma}_{+}^{m}\hat{\sigma}_{-}^{n}=\frac{1}{N}\sum_{n,m}G(k_0|\mathbf{x}_m-\mathbf{x}_n|)\sum_{\mathbf{k}'\mathbf{k}}\hat{\sigma}_{+}^{\mathbf{k}'}\hat{\sigma}_{-}^{\mathbf{k}}e^{i(\mathbf{k}\cdot \mathbf{x}_n-\mathbf{k}'\cdot \mathbf{x}_m)}\,.
\end{equation}
The summation over atoms is then split into three cases: $m>n$, $m=n$, and $m<n$, where the relations $><$ are defined on the 2D lattice, as illustrated in Fig.~\ref{fig:S1}(b) and explained in the next section. Introducing the relative coordinate $\mathbf{r}=\mathbf{x}_m-\mathbf{x}_n$ for $m>n$ (and similarly for $m<n$), the expression becomes:
\begin{align}
    \label{eq: 1exc derivation 2}
    \hat{H}_{\rm{eff}}=\frac{1}{N} \sum_{n=1}^{N}
    \sum_{\mathbf{k},\mathbf{k'}\in BZ}\sum_\mathbf{r} G(k_0|\mathbf{r}|) \,e^{i\mathbf{x}_n\cdot(\mathbf{k}-\mathbf{k}')} \Big{(}e^{-i\mathbf{k}'\cdot \mathbf{r}}+e^{i\mathbf{k}\cdot \mathbf{r}}\Big{)}\hat{\sigma}_{+}^{\mathbf{k}'}\hat{\sigma}_{-}^{\mathbf{k}}
    +G(0)\sum_{\mathbf{k},\mathbf{k}'\in BZ} \hat{\sigma}_{+}^{\mathbf{k}'}\hat{\sigma}_{-}^{\mathbf{k}}\,\cdot\frac{1}{N}\sum_{n=1}^{N}e^{i\mathbf{x}_n\cdot(\mathbf{k}-\mathbf{k}')} \,.
\end{align} 
Combining all contributions and using $\lim_{N \to \infty}\frac{1}{N}\sum_{n}e^{i\mathbf{x}_n\cdot(\mathbf{k}-\mathbf{k}')}=\delta(\mathbf{k}-\mathbf{k'})$, the final expression is:
\begin{equation}
\hat{H}_{\rm{eff}}=\sum_{\mathbf{k}}E^{(1)}_{\mathbf{k}}\hat{\sigma}_{+}^{\mathbf{k}}\hat{\sigma}_{-}^{\mathbf{k}},
\end{equation}
where the single-excitation eigenvalues read:
\begin{equation}
    \label{eq: 1exc dispersion}
E^{(1)}_{\mathbf{k}}=-i\frac{\gamma}{2}+\frac{2}{N}\sum_{\mathbf{r}}G(k_0|\mathbf{r}|)\cos (\mathbf{k}\cdot \mathbf{r})\left(\sqrt{N}-|r_x|\right)\left(\sqrt{N}-r_y\right)\, ,
\end{equation}
with the sum over $\mathbf{r}$ taking values $\mathbf{r} \in \Big{\{} \big(r_x\in \big{[}1,\sqrt{N}-1\big{]},r_y=0 \big) \Big{\}} \,\cup \, \Big{\{} \big(r_x\in \big{[}-\sqrt{N}+1,\sqrt{N}-1\big{]},r_y\in \big{[}1,\sqrt{N}-1 \big{]} \big)  \Big{\}}$. 
The real and imaginary part of Eq. \eqref{eq: 1exc dispersion} provide the the  single-excitation dispersion relation and collective decay rates, i.e. \mbox{$E^{(1)}(\mathbf k)=\epsilon^{(1)}(\mathbf k)-i\gamma^{(1)}(\mathbf k)/2$}.

\subsection{Single excitation decay rates}
In the main text, we have discussed only the thermodynamic limit for single-excitation states, where the decay rate of the eigenstates is exactly zero except for states with quasi-momenta $\mathbf{k}$ that have the same magnitude as the resonant wavevector $|\mathbf{k}|=k_0$. These states exhibit superradiant behavior. However, for finite-size systems, Bloch states $|\psi_\mathbf{k}\rangle=\frac{1}{\sqrt{N}}\sum_{\mathbf{x_m}}e^{i\mathbf{k}\cdot \mathbf{x}_m}\sigma_{+}^m|0\rangle$ are no longer exact eigenstates of the single-excitation Hamiltonian. Eigenstates of the finite-size Hamiltonian, which we label as $|\psi_s^{(1)}\rangle$, acquire finite decay rates $\gamma^{(1)}_s$. Nevertheless, to each finite-size state, we can associate a quasi-momentum $\mathbf{k}$ in such a way that $|\psi_s^{(1)}\rangle$ has the largest overlap with that Bloch state compared to all the other Bloch states. Therefore, for the rest of this section, we associate a decay rate with each quasi-momentum $\gamma^{(1)}_{\mathbf{k}}$.

\begin{figure}
    \includegraphics[width=1\linewidth]{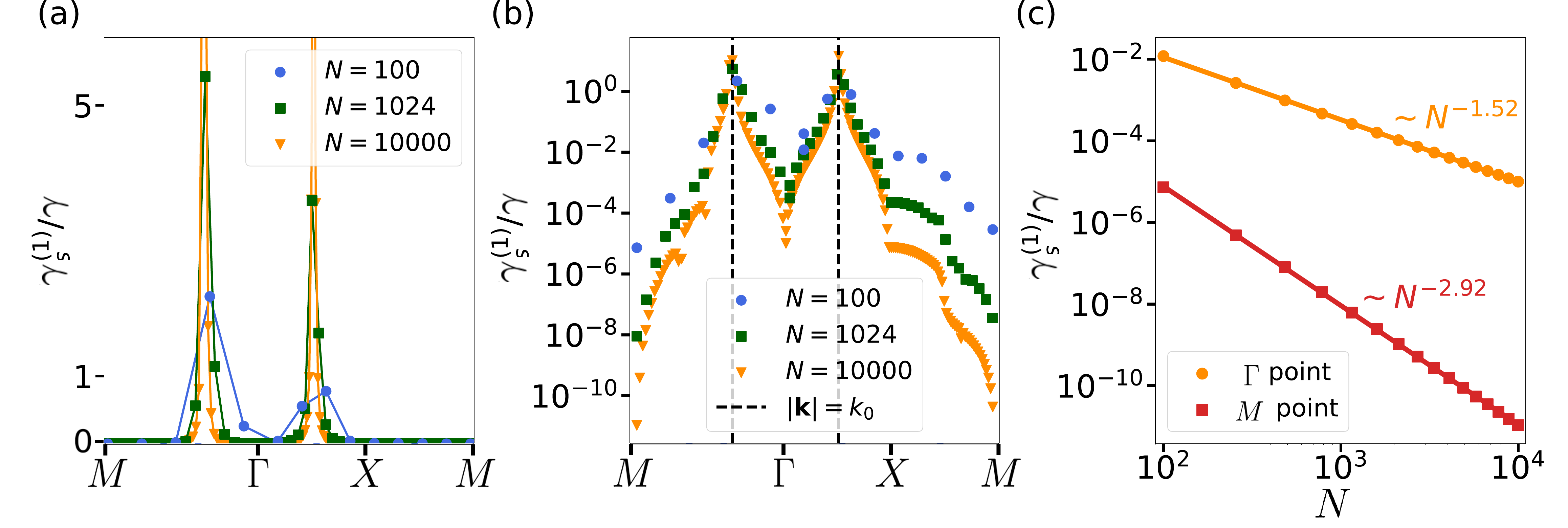}
    \caption{Decay rates of the single-excitation finite-size Hamiltonian. (a) Decay rate as a function of the unfolded Brillouin zone quasi-momenta with which the finite-size eigenstate has the largest overlap. Divergences occur for $|\mathbf{k}|=k_0$. (b) Same plot as in (a), but in logarithmic scale. (c) Scaling of the $M$ point and $\Gamma$ point decay rates with the system size. The inter-atomic distance in all plots is $k_0d=0.52\pi$.}
    
    \label{fig:S7}
\end{figure}

In Fig. \ref{fig:S7}(a) and (b), we plot the decay rate as a function of quasi-momenta in the unfolded Brillouin zone (BZ) in non-log (log) scale for different system sizes $N$. As can be observed, decay rates decrease with the system size, and the decay around the resonant wave-vector $|\mathbf{k}|=k_0$ approaches a delta function as the system size increases. Lastly, in Fig. \ref{fig:S7}(c), we study the scaling of the decay rates at the $M$ and $\Gamma$ symmetry points. We observe a power-law scaling of $N^{-2.92}$ ($N^{-1.52}$) for the $M$ ($\Gamma$) point. The latter is the one responsable for the observed decay rate of the repulsive state.

\section{Relative coordinate frame} 
In this section, we tackle the two-body problem outlined in Hamiltonian Eq.~\eqref{eq: Define general Heff} within the relative coordinate framework. Our primary goal is to find solutions for $H_{\rm eff}|\psi^{(2)}\rangle=E|\psi^{(2)}\rangle$, where the two-excitation eigenstate generally adopts the form $|\psi^{(2)}\rangle=\sum_{i<j}c^{(2)}_{ij}|e_i,e_j\rangle$, and $|e_i,e_j\rangle=\hat\sigma_+^i\hat\sigma_+^j|0\rangle$. For convenience, we re-parameterize the problem in terms of center-of-mass, $\mathbf{R}=(R_x,R_y)$, and relative, $\mathbf{r}=(r_x,r_y)$, coordinates.

In the case of an infinite system, we can take the thermodynamic limit and apply Bloch's theorem to express the center-of-mass wave function in relation to the center-of-mass momentum $\mathbf{K}$. Compared to the analogous 1D wQED case~\cite{bakkensen2021photonic,zhang2020subradiant,calajo2022emergence}, the 2D relative coordinate ansatz is more involved, and it is represented by the following expression,  schematically illustrated in Fig.\ref{fig:S1}(b),
\begin{align}
    \label{eq: ansatz}
    \begin{split}    &\big{|}\mathbf{K},\mathbf{r}\big{\rangle}=\mathcal{N}_{\mathbf{r}}\theta(r_x)\sum_{j_y=1}^{L-r_y}\sum_{j_x=1}^{L-r_x}\big{|}\phi\left(\mathbf{K},\mathbf{r},j_x,j_y \right)\big{\rangle}+\mathcal{N}_{\mathbf{r}}\theta(-r_x)(1-\delta(r_y))\sum_{j_y=1}^{L-r_y}\sum_{j_x=1+|r_x|}^{L}\big{|}\phi\left(\mathbf{K},\mathbf{r},j_x,j_y \right)\big{\rangle}, \\
    &\big{|}\phi\left(\mathbf{K},\mathbf{r},j_x,j_y \right)\big{\rangle}=e^{i\frac{K_xd(2j_x+r_x)}{2}+i\frac{K_yd(2j_y+r_y)}{2}}\big{|}(j_x,j_y),(j_x+r_x,j_y+r_y) \big{\rangle}\,.
    \end{split}
\end{align}
In the provided expression, the vector $|(i_x,i_y),(j_x,j_y)\rangle$ denotes a state where the atoms at the sites $(i_x,i_y)$ and $(j_x,j_y)$ are excited. Here, $N=L\times L$ represents the system size, and $\mathcal{N}_{\mathbf{r}}$ is the norm of the state $\big{|}\mathbf{K},\mathbf{r}\big{\rangle}$. To ensure the comprehensive counting of all two-excitation states without double counting, we assume that the relative coordinate can assume values $r_y\in [0,L-1],\,r_x\in [-(L-1),L-1]$.
Applying Eq.~\eqref{eq: Define general Heff} to the state \eqref{eq: ansatz}, we derive the following matrix elements:
\begin{equation}
\langle \mathbf K, \mathbf r| \hat{H}_\textrm{eff} | \mathbf K, \mathbf r' \rangle
    =\sum_{\epsilon=\pm 1}\cos (\mathbf{K}/2 \cdot \left(\mathbf{r}+\epsilon\mathbf{r}^{\prime}\right)) \mathcal{R}\{G\left(k_0 \left|\mathbf{r}+\epsilon\mathbf{r}^{\prime}\right|\right)\}\,,
\end{equation}
where we indicated with $\mathcal{R}\{G\left(k_0 \left|\mathbf{r}+\epsilon\mathbf{r}^{\prime}\right|\right)\}$ the real (coherent) part of the photon-mediated interaction that is the only one contributing in the thermodynamic limit.
In this manner, we have transformed the original two-body problem into a single-body problem in the relative coordinate, which depended parametrically from $\mathbf{K}$.

\section{Mapping to the impurity model}
As explained in the main text, the spin model Hamiltonian presented in Eq.~\eqref{eq: Define general Heff} can be transformed into a  hardcore boson model. This transformation involves promoting the spin operators to bosons, denoted as $\hat\sigma^i_{-}\rightarrow\hat b_i$, and introducing an on-site interaction with an infinite strength $U\rightarrow \infty$. This results in the following expression ($\hbar=1$):
\begin{equation}
    \label{eq: Hardcore bosons}
    \hat{H}^{(\rm imp)}_{\rm eff}=\hat{H}_0+\hat{H}_{\rm int}=\sum_{ij}G_{ij}\hat{b}_{i}^{\dagger}\hat{b}_{j}+\frac{U}{2}\sum_{i}(\hat{b}_{i}^{\dagger})^2\hat{b}_{i}^2,
\end{equation}
where, the bosonic operators $\hat{b}_{i}$ and $\hat{b}_{i}^{\dagger}$ respectively annihilate and create excitations on the site $\mathbf{x}_i$. In the  two-excitation sector, we can again map the problem into the center-of-mass,  $\mathbf{R}$, and relative coordinates,  $\mathbf{r}$, with  the center-of-mass momentum $\mathbf{K}$ being a well-defined quantum number in the thermodynamic limit. 
The first term in Hamiltonian~\eqref{eq: Hardcore bosons} accounts for free bosons, and its energy is straightforwardly the sum of individual photon energies. This term can be expressed in a diagonal form with respect to the relative momentum state $|\mathbf K,\mathbf q\rangle$.
\begin{equation}
  \hat H_{0}= \sum_{\mathbf q }\epsilon^{(2)}_{\text{scat}}(\mathbf{K},\mathbf q)|\mathbf K,\mathbf q\rangle\langle \mathbf K, \mathbf q|
\end{equation}
where  $\epsilon^{(2)}_{\text{scat}}(\mathbf{K},\mathbf q)=\epsilon^{(1)}(\mathbf{q})+\epsilon^{(1)}(\mathbf{K}-\mathbf{q})$. We can Fourier transform this Hamiltonian with respect to the relative coordinate using the relation $|\mathbf K,\mathbf q\rangle=(1/\sqrt{N})\sum_{\mathbf r }e^{i\mathbf q\mathbf r}|\mathbf K,\mathbf r\rangle$. By incorporating the interaction term mapped to the relative coordinate space, we arrive at the impurity model Hamiltonian presented in the main text.
 \begin{equation}
    \label{eq: impurity H}
    \hat H_\mathbf{K}=\sum_{\mathbf r \mathbf r'}J^{\mathbf K}_{\mathbf r \mathbf r'}|\mathbf K, \mathbf r \rangle\langle \mathbf K, \mathbf r'|+U|\mathbf K, 0 \rangle\langle \mathbf K, 0|,
\end{equation}
where $J^{\mathbf K}_{\mathbf r \mathbf r'}=(1/N)\sum_{\mathbf{q}}\omega_{\mathbf{K}}(\mathbf{q})e^{i(\mathbf r -\mathbf r')\mathbf{q}}$ and $|\mathbf K, 0\rangle=|\mathbf K,\bold{r}=0\rangle$ is the site where is placed the impurity. 

As discussed in~\cite{economou2006green}, it is possible to derive a closed expression for the single-particle Green function in this particular problem:
\begin{equation}
\mathcal{G}=\mathcal{G}_0+\mathcal{G}_0|\mathbf K,0\rangle\frac{U}{1-U\mathcal{G}_0(0,E)}\langle \mathbf K,0|\mathcal{G}_0\,,
\end{equation}
where $\mathcal{G}_0$ is the Green function of the non interacting Hamiltonian $\hat H_0$ and 
\begin{equation}
    \mathcal{G}_0(\mathbf r,E)=\langle \mathbf K,\mathbf r|\mathcal{G}_0(E)|\mathbf K,\mathbf 0\rangle=\frac{1}{(2\pi)^2}\int_{\rm BZ} d^2\bold{q}\frac{e^{i\bold{q}\cdot\bold{r}}}{E-\epsilon^{(2)}_{\text{scat}}(\mathbf{K},\mathbf q)}\,.
\end{equation}
The poles of the Green function, characterized by energies $E_{\rm BS}$, correspond to the bound state solutions of the system. The equation governing these bound states is expressed as:
\begin{equation}
\mathcal{G}_0(0,0,E_{\rm BS})=\frac{1}{U}\rightarrow 0^+\,.
\end{equation}
This equation can be numerically solved by seeking solutions in the gap for each $\bold{K}$ value:
\begin{equation}
\sum_{\bold{q}\in _{\rm BZ}}\frac{1}{E_{\rm BS}-\epsilon^{(2)}_{\text{scat}}(\mathbf{K},\mathbf q)}= 0\,,
\end{equation}
where the sum over the relative momentum spans the Brillouin zone.
Finally with the Green's function formalism, we can also express the solution for the unnormalized bound states in the relative coordinate:
\begin{equation}\label{state}
|\psi_{\rm BS},\mathbf{K}\rangle=\sum_\bold{r}\mathcal{G}_0(\bold{r},E_{\rm BS})|\mathbf{K},\bold{r}\rangle.
\end{equation}


\begin{figure}
    \includegraphics[width=0.99\linewidth]{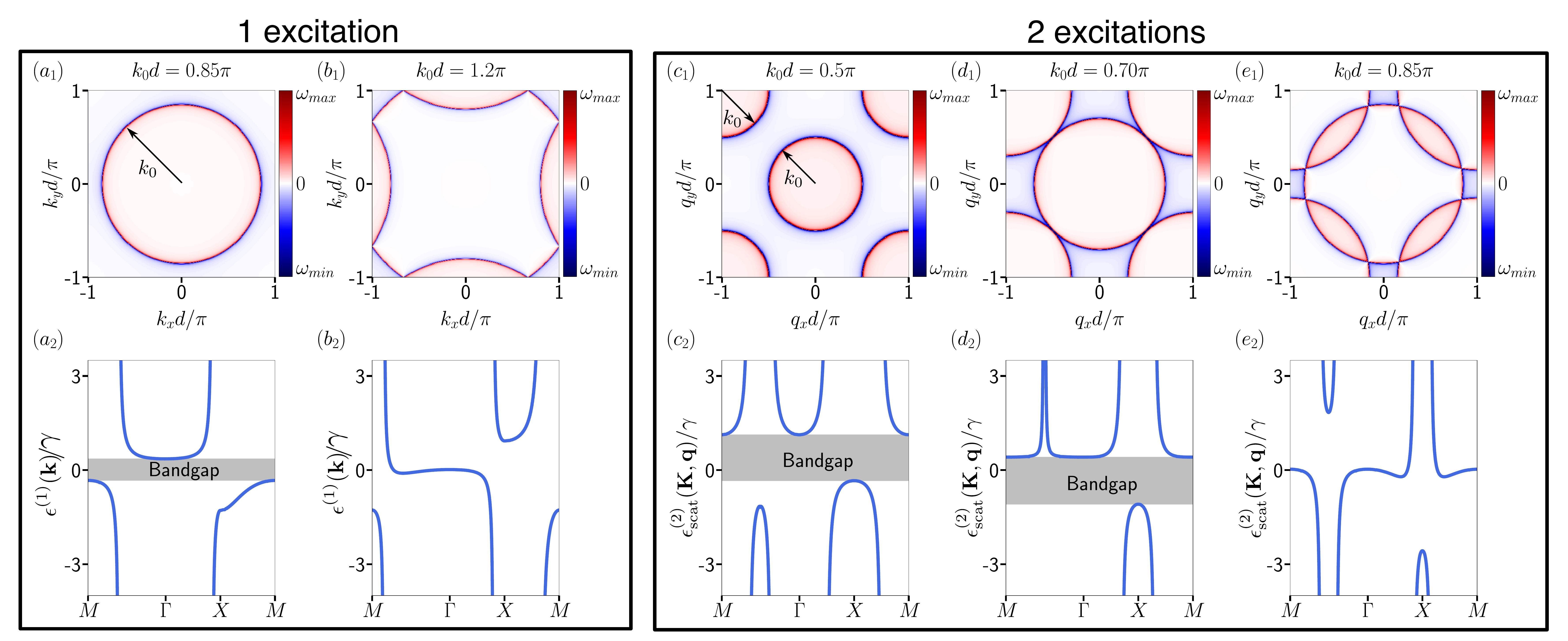}
    \caption{(a)-(b)Single photon dispersion  relation  at various inter-atomic distances both before and after the gap closing at $k_0d=\pi$. Panels (c)-(e) illustrate the two-excitation dispersion of unbound photons, denoted as $\epsilon^{(2)}_{\text{scat}}(\mathbf{K},\mathbf q)$, as a function of the relative momentum $\mathbf{q}$ with the center-of-mass momentum fixed at the $M$-point, $\mathbf{K}=\left(\frac{\pi}{d},\frac{\pi}{d}\right)$.
In the first row (1), the dispersions are plotted within the first Brillouin zone, while in the second row (2), they are unfolded with respect to the $\Gamma$, $X$, and $M$ symmetry points.
All plots are generated for a system size of $N=300\times 300$.     }
    \label{fig:S2}
\end{figure}

\section{Bandgap closing}
As explained in the main text, the energy of bound states, denoted as $E_{\rm BS}$, must differ from the sum of individual photon energies, given by: 
\begin{equation}\label{Eq.exist_Bs}
    E_{\rm BS}(\mathbf{K})\neq\epsilon^{(1)}(\mathbf{q})+\epsilon^{(1)}(\mathbf{K}-\mathbf{q})\,.
\end{equation}
This condition cannot be satisfied in the absence of a gap in the single excitation spectrum. In Fig.1 of the main text, we demonstrated the closing of the single excitation bandgap above the sub-wavelength limit, $k_0d>\pi$. This closing is visibly apparent by examining the single-particle dispersion at different inter-atomic distances, as illustrated in Fig.~\ref{fig:S2}(a)-(b). When $k_0d>\pi$, the central polaritonic branch with a radius of $|\mathbf{k}|=k_0$ reaches the edges of the 1BZ, and higher diffraction orders prevent gap existence.
Given that the condition given in Eq.~\eqref{Eq.exist_Bs} cannot be satisfied for $k_0d>\pi$, we focus on studying the unbound part of the two-excitation spectrum
$\epsilon^{(2)}_{\text{scat}}(\mathbf{K},\mathbf q)=\epsilon^{(1)}(\mathbf{q})+\epsilon^{(1)}(\mathbf{K}-\mathbf{q})$ for different values of $k_0d<\pi$ as a function of the relative momentum $\mathbf{q}$. This allows us to determine the regions of validity for~\eqref{Eq.exist_Bs}.
At the $\Gamma$ point ($\mathbf{K}=\left(0,0\right)$), the gap closes precisely at the same distance as for a single excitation because trivially $\epsilon^{(2)}_{\text{scat}}(\mathbf{K},\mathbf q)=2\epsilon^{(1)}({\mathbf{q}})$.
At the $M$ point ($\mathbf{K}=\left(\frac{\pi}{d},\frac{\pi}{d}\right)$), the dispersion reveals multiple polaritonic branches with a radius of $|\mathbf{q}|=k_0$, positioned in the middle and at the edges of the Brillouin zone. The bandgap then closes when these polaritonic branches intersect for $k_0d\geq \frac{\pi}{\sqrt{2}}$, as illustrated in Fig.~\ref{fig:S2}.

\begin{figure}
\includegraphics[width=0.75\linewidth]{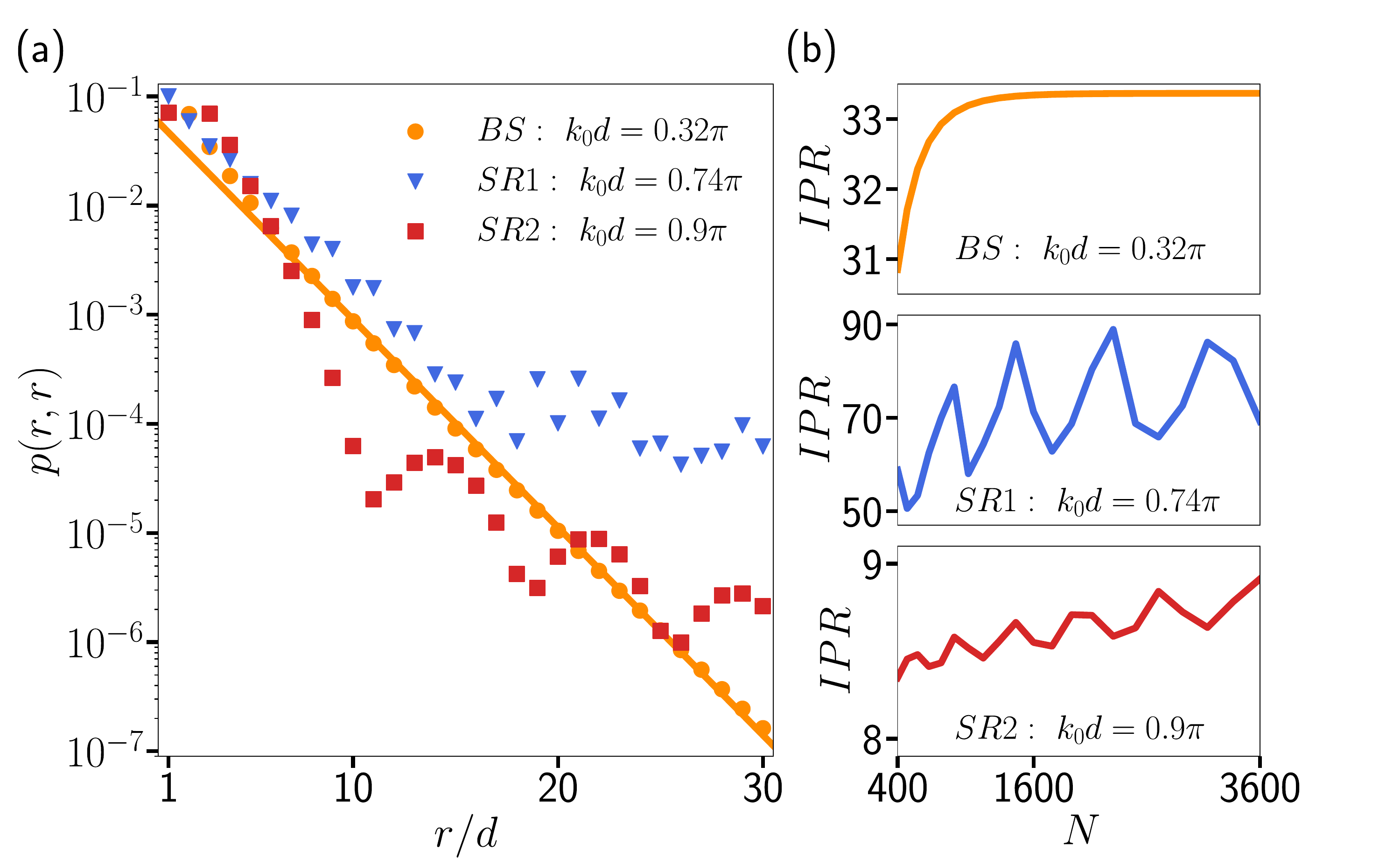}
    \caption{ (a) The distribution of relative coordinates for $r_x=r_y=r$ is depicted for the $M$ point bound state (BS), marked by orange dots, along with two scattering resonances (SR) at the $M$ point. SR1, represented by blue triangles, denotes the scattering resonance found on the same energy branch as the bound state when the bandgap closes.  SR2, indicated by red squares, represents the scattering resonance from the other energy branch located in the low density of states region of the 2-excitation dispersion spectrum (see Fig. 2 of the main text).
(b) The inverse participatio ratio is shown as a function of system size for the same states as in panel (a). In all the plots we considered a system size of $N = 60 \times 60$ }
    \label{fig:S3}
\end{figure}

 \section{In-band interacting states}
While the bound states discussed in the main text lie in the band-gap and can be easily discerned by their isolated energies,  the repulsive states and scattering resonances exist within the continuum and cannot be identified by spectral analysis.
To pinpoint the repulsive states, we draw upon insights from the analogous case of 1D waveguide QED~\cite{sheremet2023waveguide}, where it is established that these states manifest as highly subradiant. Accordingly, we seek them among the two-excitation states with the lowest decay rates (in a finite size system).

The scattering resonances, even if embedded in the band, exhibit localization of the two excitations with respect to their relative coordinates. This arises from the fact that, by definition~\cite{RevModPhys.66.539,bakkensen2021photonic}, they are made by a superposition of a bound and an unbound component. We utilize this property to identify these states via the inverse participation ratio (IPR), defined as:
\begin{equation}
  IPR=\frac{1}{\sum_{\mathbf{r}}\big{|}\psi(\mathbf{r})\big{|}^4}\,,
\end{equation}
which allow us to filter the most localized states out of the spectrum.\\

 Even if localized,  these states are quantitatively distinct from bound states. This distinction is evident in Fig.~\ref{fig:S3}(a), where we plotted the spread of the relative coordinate population distribution for $r_x=r_y$ for the two scattering resonance branches discussed in the main text.
The plot illustrates that bound states exhibit an exponential decay with the relative distance, whereas scattering resonances display a non-exponentially oscillating tail. Note that a similar behavior
has  been discussed also for 1D waveguides~\cite{bakkensen2021photonic}.
 To further verify the nature of these states, in Fig.~\ref{fig:S3}(b) we plotted the inverse participation ratio 
 as a function of the system size. As observed, for bound states at $k_0d=0.32\pi$, the IPR saturates to a constant value for $L\gtrapprox 40$. Conversely, for scattering resonances at $k_0d=0.74\pi,0.9\pi$, the IPR oscillates without reaching an asymptotic value, thereby confirming the bound and scattering resonance nature of these states, respectively.


\begin{figure}
    \includegraphics[width=1.0\linewidth]{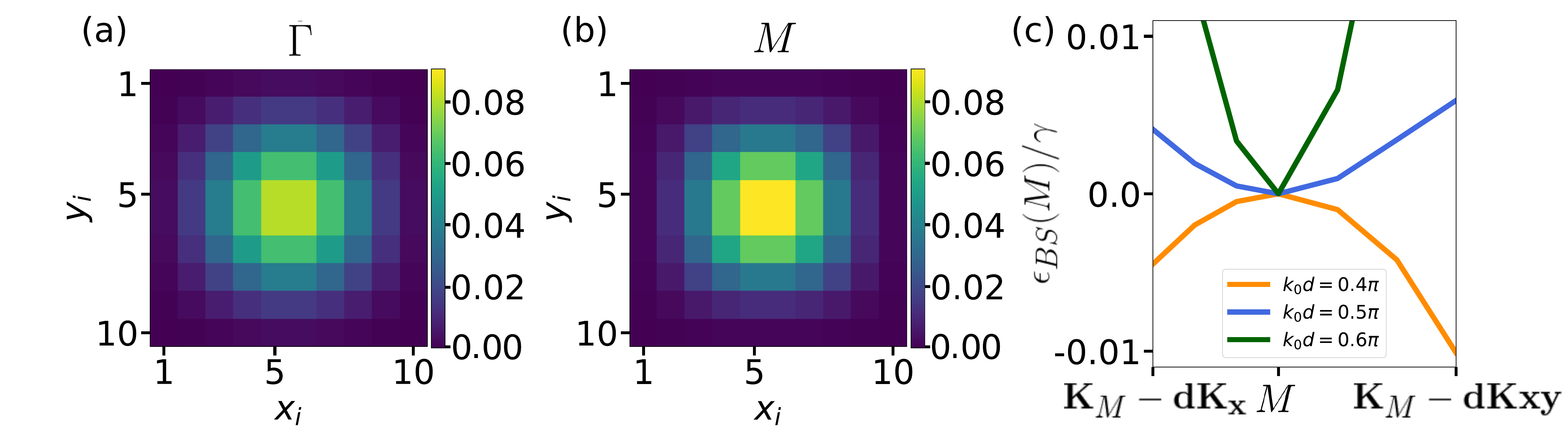}
    
    \caption{(a),(b) panels show the real space excitation distribution for the $\Gamma$, $M$ point BS, respectively. Results were obtained by diagonalizing full two-excitation Hamiltonian for the system size $N=10\times 10$ at the inter-atomic distance of $k_0d=0.5\pi$. (b) Bound state dispersion relation as a function of center-of-mass momentum $\mathbf{K}$ in the vicinity of M symmetry point for different interatomic distances. The dispersion is plotted for the unfolded region spanning from $\mathbf{K}_M-\mathbf{dK_x}$ to $\mathbf{K}_M$ along the X-M symmetry point line and then to $\mathbf{K}_M-\mathbf{dK_{xy}}$ along the M-$\Gamma$ symmetry point line, where we have defined $\mathbf{K}_M=(1,1)\frac{\pi}{d}, \,\mathbf{dK_x}=(0,0.1)\frac{\pi}{d}, \,\mathbf{dK_{xy}}=(0.1,0.1)\frac{\pi}{d}$. Results were obtained by diagonalizing relative coordinate Hamiltonian for the system size $N=40\times 40$.
    }  \label{Fig:S8} 
   
\end{figure}

 \section{Bound states localization in the center of mass coordinate}
To explain the nontrivial dependence of the M point bound state lifetime as a function of inter-atomic distance, it is not sufficient to look just at the localization of these states in the relative coordinate; it is also important to consider their localization in the center of mass coordinate. In particular, the $M$ point BS results in being significantly more long-lived compared to the $\Gamma$ point.
To illustrate this behaviour, in Fig.~\ref{Fig:S8}(a), we plot the population distribution of the $\Gamma$ and $M$ point BSs at the inter-atomic distance of $k_0d=0.5\pi$, where the states are more sub-radiant. It can be observed that the latter is more localized in the center of the bulk, resulting in weaker leaking of the photons from the system.
Such localization can be explained by examining the dispersion as a function of the center-of-mass momentum, $\mathbf{K}$, around the $M$ point, as shown in  Fig.~\ref{Fig:S8}(b). The dispersion is flattest close to the inter-atomic distance $k_0d=0.5\pi$, leading to a larger effective mass for the two-paired excitations, which take more time to leak through the edges (see Ref.~\cite{poddubny2020quasiflat} for a similar effect in 1D wQED).


\begin{figure}
    \includegraphics[width=1.05\linewidth]{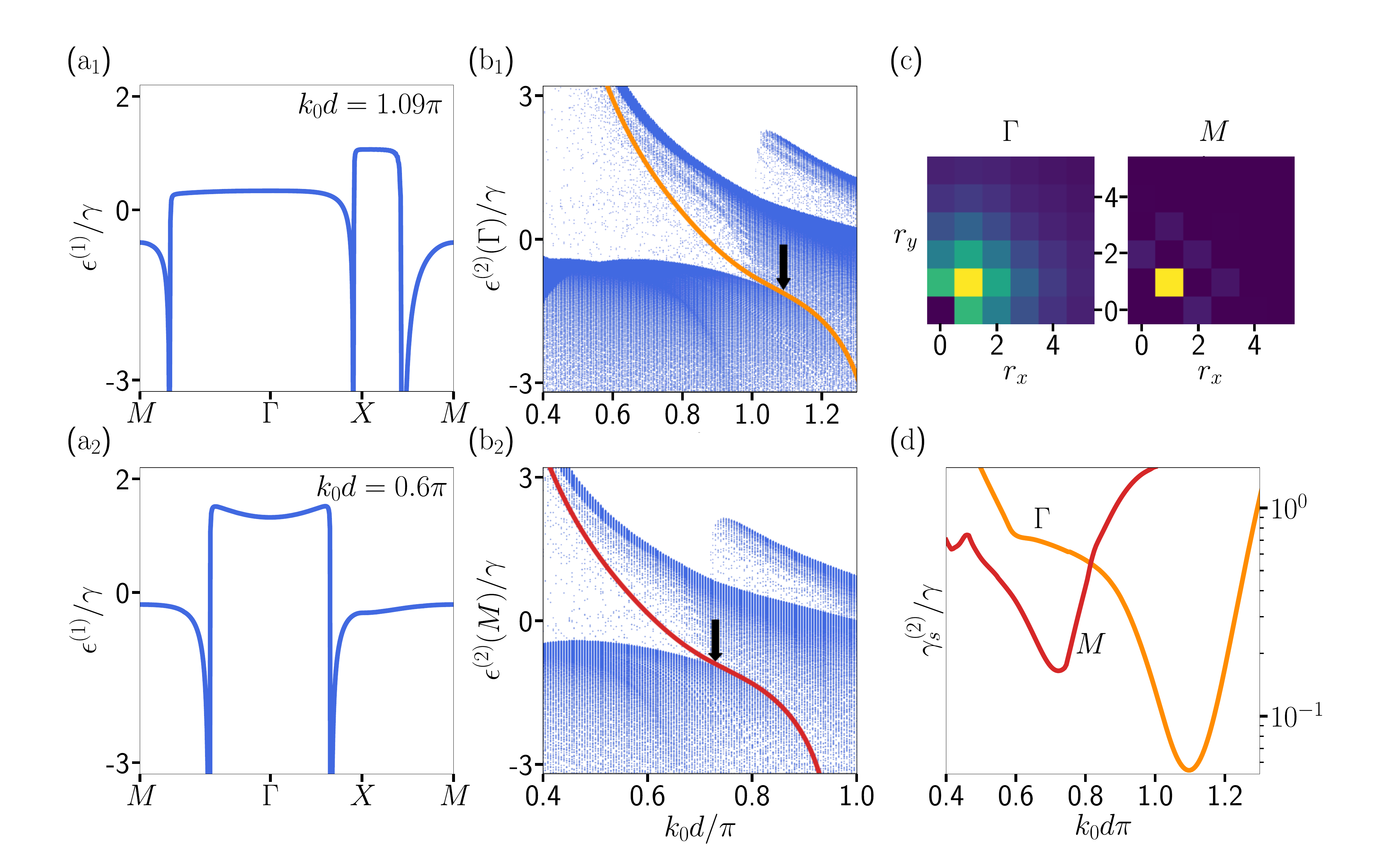}
    \caption{2D free space array with atomic polarization orthogonal to the array plane. $\mathrm{(a_1),(a_2)}$ Single excitation dispersion for $k_0d=1.09\pi,\,0.6\pi$. $\mathrm{(b_1),(b_2)}$ Two excitation spectrum at the  $\Gamma$ and $M$ point of center-of-mass momenta as a function of inter-atomic distance. Blue dots represent the sum of the free photon energies, while the red (orange) line represents the energy of the scattering resonance. (c) Probability distribution of the  $\Gamma$  and $M$ point scattering resonance  for $k_0d=0.6\pi,\,1.09\pi$. (d) Decay rate of $\Gamma$ and $M$ point scattering resonances as a function of $k_0d$ for system size $N=10\times 10$. In panels (a),(b), we considered a system size of $150 \times 150$, while in panel (c) we considered a system size $60 \times 60$.  
    }  
    \label{fig:S5}
\end{figure}

\section{Interacting states in  2D free space atomic arrays}
 The dyadic Green's tensor, which characterizes the interaction between dipoles in free space  employed in the main text, is derived as a solution to the dyadic equation \cite{lehmberg1970radiation,dung1998three,asenjo2017atom,perczel2017topological}:
\begin{equation}
    \label{eq: dyadic equation}
    \frac{\omega_A^2}{c^2}\tilde{G}_{\alpha \beta}(\mathbf{r},\mathbf{r}')-\left(\partial_{\alpha}\partial_{\nu}-\delta_{\alpha \nu}\partial_{\eta}\partial_{\eta} \right) \tilde{G}_{\nu \beta}(\mathbf{r},\mathbf{r}')=\delta_{\alpha\beta}\delta(\mathbf{r}-\mathbf{r}')\,.
\end{equation}
where the Green's tensor, denoted as $\tilde{G}_{\alpha \beta}(\mathbf{r},\mathbf{r}')$, characterizes the component $\alpha={x,y,z}$ of the electromagnetic field at position $\mathbf{r}$ emitted by a perfect dipole positioned at $\mathbf{r}'$ and oriented in the $\beta={x,y,z}$ direction. The equation solution is explicitly given by:
\begin{equation}
    \label{eq: Dyadic Green}
    \tilde{G}_{\alpha \beta}(\mathbf{r}-\mathbf{r}')=\frac{e^{ik_0r } }{4\pi k_0^2r^3}\Bigg{[}\left(1-i(k_0r)-(k_0r)^2 \right)\delta_{\alpha \beta}+\left(1+\frac{3i}{k_0r}-\frac{3}{(k_0r)^2} \right)k_0^2r_\alpha r_\beta \Bigg{]},
\end{equation}
where we have defined $r=|\mathbf{r}-\mathbf{r}'|$, the resonant wave vector $k_0=\frac{\omega_a}{c}$, and with $r_\alpha$ ($\alpha=\{x,y,z\}$) we indicated the different components of the vector $\mathbf{r}-\mathbf{r}'$ in the Cartesian coordinates. In the main text, we configure the atomic polarization to be oriented perpendicular to the array's plane. In this particular orientation, it is sufficient to consider only the $\tilde{G}{zz}$ component of the Green's tensor. This component yields the collective interaction function used in the main text, denoted as $G(x)=\frac{3\pi\gamma_0}{k_0} \tilde{G}{zz}$ with $\gamma_0=d_{eg}^2k_0^3/(3\pi\hbar\epsilon_0)$ being the free space spontaneous emission decay rate and $d_{eg}$  the dipole moment strength.
Once identified the collective interaction term let's now discuss the single and two excitation spectrum.
In contrast to the 2D confined scenario,  the single excitation spectrum of the free space array lacks a bandgap, as shown in Fig.~\ref{fig:S5}(a).
This absence of a bandgap extends to the two-excitation spectrum, implying the nonexistence of bound states in this system. 
Nonetheless, subradiant repulsive and scattering resonances (SRs) persist in this setting within the subradiant region enclosed in the first BZ, $k_0d<\sqrt{2}\pi$. The former can be identified in a manner similar to the 2D confined case and exhibit a comparable probability distribution of the two excitations. The latter are more involved, as depicted in Fig.~\ref{fig:S5}(b), where we illustrate the two-excitation spectrum for the $\Gamma$ and $M$  points, highlighting the SRs branches. 
Similar to the confined case, the existence of spectral regions with a low density of states can result in highly bound SRs, as demonstrated in Fig.~\ref{fig:S5}(c) through the relative coordinate probability distribution of the two SRs.
Importantly, these SRs exhibit subradiant behavior even if they are composed of two attractive excitations that can potentially scatter to each other in free space. Their decay rate is strongly dependent on the inter-atomic distances, as depicted in Fig.~\ref{fig:S5}(d), showing pronounced minima, respectively at $k_0d=1.09\pi$ and $k_0d=0.73\pi$ (highlighted in Fig.~\ref{fig:S5}(b) by the black arrow), where it becomes more than ten times smaller than the single atom emission rate. Importantly, both critical distances are moderately subwavelength making this regime potentially accessible  in cold atom arrays~\cite{rui2020subradiant}. 

Note that the  behaviour of the SR at the $\Gamma$ point is more subtle. Indeed,  this point is known to host subradiant states in single excitation only for orthogonal atomic polarization. In this setting, all the atoms oscillate in phase and  orthogonally  to the array, limiting the emission strictly within that plane~\cite{asenjo2017atom}.
Considering this, the extended lifetime observed in the scattering resonance at the $\Gamma$ point appears to be connected to the remarkably flat dispersion at this point (see Fig.~\ref{fig:S5} $\mathrm{(a_1)}$). This flat dispersion  induces localization in the center of mass wavefunction, leading to a reduction in emissions at the edges (see~\cite{poddubny2020quasiflat}).

\begin{figure}
    \includegraphics[width=1\linewidth]{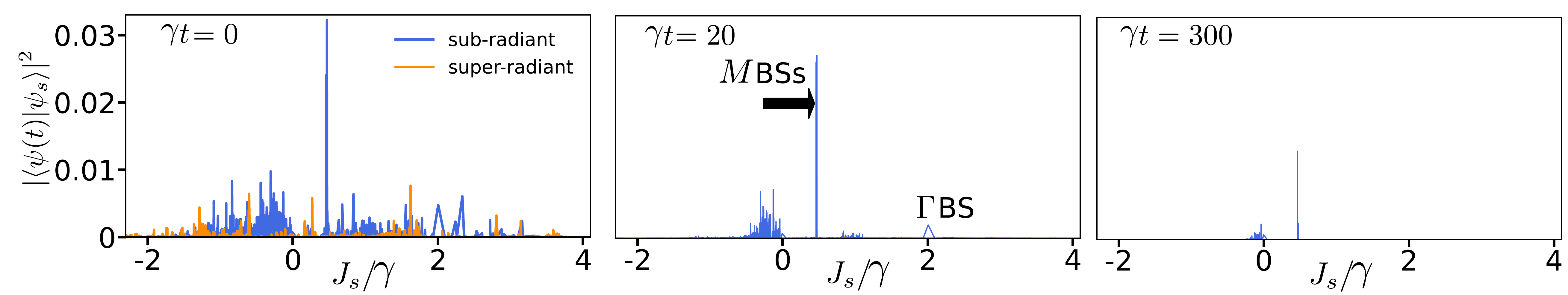}
    \caption{(a) The projection of the state $|\psi(t)\rangle$ onto the eigenstates in the two-excitation sector is displayed at times $\gamma t=0,20,300$. The most prominent projection comes from the states with center-of-mass momentum close to the $M$ point (labeled by $M$ BSs in the second panel). The peak at $J_s\approx 2$ corresponds to the $\Gamma$ point bound state, while the group of peaks at $J_s\approx 0$ corresponds to other long-lived states.
In all plots, the parameter is set as $k_0d=0.52\pi$, and an array size of $N=10\times 10$ is considered.
    }  
    \label{fig:S4}
\end{figure}

\section{Dynamical excitation of the bound states}
As detailed in the main text, we investigate the dynamical excitation of bound states by solving the time-dependent Schr\"odinger equation governed by the effective Hamiltonian~\eqref{eq: Define general Heff},  projected in the double excitation sector ($\hbar=1$):
\begin{equation}\label{eq:time_Shro}
 i\frac{\partial}{\partial t}|\psi^{(2)}(t)\rangle= H_{\rm eff}|\psi^{(2)}(t)\rangle\,.
\end{equation}
In this context, the initial state is set as 
\mbox{$|\psi(t=0)\rangle=\hat\sigma_+^\mathbf{i}\hat\sigma_+^\mathbf{i+\boldsymbol{\ell}}|0\rangle$},
where the initially excited atoms are separated by a distance $\mathbf{x}_i-\mathbf{x}_j=\boldsymbol{\ell}=(\ell,\ell)$. The evolution of the state according to \eqref{eq:time_Shro} neglects the recycling term in the master equation \eqref{eq: Master equation}. Consequently, the norm of the state is not conserved in time, $\big{|}\langle \psi(t) | \psi(t)\rangle\big{|}^2<1$, as the population of the single and zero excitation subspaces is not monitored during the state relaxations. This approach yields accurate results as long as observables are computed within the two-excitation subspace.

In the main text, we examine the excitation of the M point bound state starting from an initial state with $\ell=1$, where two next-nearest neighbor atoms (along the diagonal of the square plaquet) are excited. This initial condition, when projected onto the (incomplete) two-excitation sector non-Hermitian basis, significantly overlaps with the bound states having a center-of-mass momentum $\mathbf{K}$ close to the $M$ point, as illustrated in the first panel of Fig.~\ref{fig:S4}(a). In this figure, an initial non-zero projection with other sub-radiant (in blue) and super-radiant states (in orange) is also observed. As the state evolves (second and third panel of Fig.~\ref{fig:S4}(a)), contributions from other states drop, while those from the $M$ point bound states persist over time. A weak excitation of other long-living states is also noted. This includes the $\Gamma$ point bound state (less subradiant than the M point) located at energy $J_s\approx 2$, where $s$ is an index labeling the states and a contribution from other subradiant states located at $J_s\approx 0$.

\section{Experimental considerations}
In this section, we discuss possible experimental implementations of the model proposed in the main text within the current state-of-the-art experiments.

\subsection{Optical domain }

In the optical domain, one of the most prominent ways of realizing the model proposed in this letter is to couple a 2D atomic array to a 2D photonic crystal waveguide (PCW) with specifically tailored dispersion. Specifically, the atomic transition frequency can be set near the approximately isotropic quadratic band edge (see Sec.~\ref{Sec.spinmodel}) of a PCW, which usually occurs in the vicinity of high symmetry points.

In Ref.~\cite{gonzalez2015subwavelength}, an extensive numerical study was performed on a realistic 2D wQED setup based on Gallium Phosphide (GaP) with a refractive index of $n=3.25$. The proposed photonic crystal waveguide (PCW) was tailored to trap via Casimir-Polder forces $^{87}\text{Rb}$ atoms in a sub-wavelength lattice and utilize its $D_2$ transition at a wavelength of $\lambda_0=770, \mathrm{nm}$ for photon-mediated atom-atom interactions. With this scheme it should be possible to trap the atoms at a distance of approximately $d=50\,\mathrm{nm}$.
At such a distance, the decay rate into the 2D structure was estimated to be $\frac{\gamma}{2\pi}\approx 1-10^{3},\mathrm{MHz}$, which can be significantly larger than the free-space decay rate of the $D_2$ line of $^{87}\text{Rb}$ atoms, $\frac{\gamma_0}{2\pi}=6.07 \mathrm{MHz}$.

The first experimental attempts toward this scenario have been pursued in Ref.~\cite{yu2019two}, where a 200 nm thick photonic crystal waveguide  based on Silicon nitride (SiN), a low-loss material with $n=2$, was fabricated. The PCW was designed to trap Cs atoms and couple the band edge to the Cs D1 transition at the vacuum wavelength $\lambda_0=894\,\mathrm{nm}$.

In this section we concentrated on cold atoms coupled to PCWs but another experimentally relevant possibility relies in considering artificial atoms such as quantum dots~\cite{lodahl2015interfacing} and diamond defects~\cite{sipahigil2016integrated} coupled to integrated photonic structures.
In all the mentioned experimental settings, the most significant hurdles consist of achieving efficient coupling to the structure and preparing an ordered array of emitters, which involves trapping challenges for cold atom implementation and control over defect positions for artificial atoms. These limitations can be highly relaxed in the microwave domain that instead comes with scalability challenges, as discussed in the following section.

\subsection{Microwave resonator arrays}
In recent years, significant progress has been made in 
coupling superconducting qubits arrays either to transmission lines~\cite{brehm2021waveguide} or large and scalable microwave resonator arrays~\cite{zhang2023superconducting,scigliuzzo2022controlling}.
In Ref.~\cite{brehm2021waveguide}, it was demonstrated the coupling and control of eight transmon qubit, with frequency $\omega_a\approx 3-8\,\mathrm{GHz}$, coupled to a one-dimensional transmission line.
The individual qubit decay rate to the waveguide was reported to be $\frac{\gamma_{1D}}{2\pi}= 6.4,\mathrm{MHz}$, which is high compared to the observed nonradiative decays of $\frac{\gamma_{nr}}{2\pi}= 240-560,\mathrm{KHz}$. Even larger qubit-resonator couplings were obtained in Ref.~\cite{scigliuzzo2022controlling}, where two transmon qubits ($\omega_a\approx 6.5,\mathrm{GHz}$) were coupled to an array of microwave resonators composed of Josephson junctions. Qubit-array coupling was reported to be as strong as $g\approx 340 ,\mathrm{MHz}$, which resulted in efficient qubit-qubit coupling of $\sim 50,\mathrm{MHz}$. With such strong couplings, microwave waveguides coupled to superconducting qubits are strong candidates for studying collective 2D wQED effects.
Major challenges in scaling these settings toward 2D architecture rely on cross talk and impurious coupling both among the emitters and the resonators, which may induce disorder in the system.

\subsection{Free space sub-wavelength atomic arrays}

Despite the significant experimental progress, efficiently coupling sub-wavelength atomic arrays to 2D photonic structures remains an ongoing challenge. On the other hand, in the main text, we demonstrated that some of the strongly interacting photon-photon states that have been discussed can also be observed in free space sub-wavelength atomic arrays. Large sub-wavelength 2D free space arrays have already been implemented in state-of-the-art experiments. In Refs.~\cite{rui2020subradiant} and ~\cite{srakaew2023subwavelength}, arrays of, respectively, $N\sim 200$ and $N\sim 1500$ $^{87}\text{Rb}$ atoms were realized in a sub-wavelength optical lattice with a lattice spacing of $d=532,\mathrm{nm}$. Linear optical responses have been studied experimentally, showing that the array can lead to a cooperative response acting as an atomic mirror to orthogonally incident light. This collective regime is the same where the finding of our study could be promptly probed.

\end{document}